\documentstyle[course,epsf,rotate]{houches}

\newcommand{\r}{{\bf r}}
\newcommand{\q}{{\bf q}}

\newcommand{\ep}{\epsilon}

\makeatletter 
\def\@copyright{\vfill \noindent
  \parbox{0.75\textwidth}{\raggedright \small\em
  to be published in:  \\
  Les Houches, Session LXIII, 1995\\
  E. Giacobino, S. Reynaud and J. Zinn-Justin, eds. \\
  Fluctuations Quantiques  \\
  Quantum Fluctuations       \\
  Elsevier Science}}
\makeatother
\newcommand{\gtequiv}{\lower2pt\hbox{$\:\stackrel{>}{
\scriptstyle\sim}\:$}}
\newcommand{\ltequiv}{\lower2pt\hbox{$\:\stackrel{<}{
\scriptstyle\sim}\:$}}
\begin{document}
\author[G. Montambaux]{G. Montambaux}
\address{Laboratoire de Physique des Solides, Associ\'e au CNRS    \strut\\
	Universit\'e Paris--Sud       \strut\\
	 91405 -- Orsay            \strut\\
	 France           \strut\\
		    \strut\\
	 February 5th, 1996    \strut \\
	 corrections  Feb. 1997    \strut \\  }
\chapter[Spectral Fluctuations in  Disordered
Metals
]{Spectral fluctuations \\ in disordered metals
}

\section{Introduction, the relevant scales} \label{sect1}
With the development of the nanotechnologies, new
possibilities recently emerged to find electronic
properties which reflect the discrete aspect of energy spectra. 
The objects  we shall describe in this course are metals
which will be considered as  quantum mechanical objects.
Long time ago Kubo realized   that in a small system of
typical size $L$, the inter-level
spacing $\Delta \propto L^{-d}$ may become as large as the
 temperature.
The discreteness of the spectrum should  manifest itself
through a qualitative change in thermodynamic
properties\cite{Kubo62,Brody81}. It turns out that beside $\Delta$, another
characteristic energy scale $E_c$ emerges  ({\it the Thouless energy}),
 which  determines the thermodynamic properties in a crucial way.

 It is the purpose of this course to review the properties of the metallic
spectra and relate them to physical quantities, transport and
thermodynamics.  In a first approximation, a metal can be considered as
a complex quantum system which shares universal spectral properties
 with other so-called chaotic systems, like nuclei, molecules, models
of billiards. One essential property is that the energy levels are
strongly correlated and present the phenomenon of {\it spectral rigidity}
\cite{Brody81}.

Such a  system is described by four
length
scales: the sample size $L$, the mean free path $l_e$
which
describes the elastic collisions, the Fermi wave length $\lambda_F$ which
depends on the density of electrons and the coherence length $L_\varphi$.
This last scale is very important  because the effects  we aim
 to describe in this
course result from the phase coherence of the wave functions and thus
disappear beyond $L_\varphi$\cite{Lphi}. Smaller distances define {\it
the mesoscopic regime}.
In this course, we shall mostly consider an electron gas in the
following limits:
\begin{equation}
\lambda_F \ll l_e \ll L \ll L_\varphi
\label{Escales}
\end{equation}
which correspond to a {\it weakly disordered} ($\lambda_F \ll l_e$),
{\it mesoscopic} ($L \ll L_\varphi$) metal
in the {\it diffusive regime} ($l_e \ll L$).
When the disorder becomes so large that $l_e$ is reduced to a length
of order  $\lambda_F$, the wave functions become localized on a typical
scale $\xi$ called the localization length\cite{Anderson58,Abrahams79}. Here
we shall study only the diffusive regime where
$\xi \rightarrow \infty$.

In this regime, an
electron moves diffusively because it experiences many
elastic collisions while moving in the sample. The
typical distance covered by the diffusive particle in a time $t$ varies as
$r^2(t) \simeq D t$. The diffusive motion is thus characterized by a new
time scale $\tau_D$ which is the typical time for an electron to
travel through the sample. It is defined
 as $\tau_D = L^2/D$.   To this
time scale corresponds a new energy scale $E_c$ called the
Thouless energy:
\begin{equation}
E_c = {\hbar \over \tau_D} = {\hbar D \over L^2}
\label{thouless}
\end{equation}
where the classical diffusion coefficient $D$ is  given by
\begin{equation}
D ={v_F^2 \tau_e \over d}={v_F l_e \over d}
\label{diffusion}
\end{equation}
$\tau_e = l_e/v_F$ is the elastic collision time. $v_F$ is the Fermi velocity.
It is interesting to notice
that the d.c. residual ($T=0K$) conductance
$\sigma$ can be directly related to the Thouless energy.
Using the Einstein relation $\sigma = e^2 \rho_0 D$, where $\rho_0$ is the
average density of states, and the Ohm relation between the conductance $G$
and the conductivity, $G = \sigma L^{d-2}$, the
dimensionless conductance $g$ can be written as
 \begin{equation}
 g={ G \over e^2/\hbar}  ={\sigma  L^{d-2}  \over e^2/\hbar}=
{E_c \over \Delta} \equiv N(E_c)\label{g}
\end{equation}
since the average interlevel distance $\Delta$ is $1/(\rho_0
L^d)$. $N(E_c)$ is thus the average number of states in a strip of the
spectrum
of width $E_c$. We shall see that, beside $\Delta$, the energy scale
$E_c$  enters in an essential way to describe
the spectral properties of mesoscopic systems.
\begin{figure}[hb]
\centerline{
\epsfxsize 10cm
\epsffile{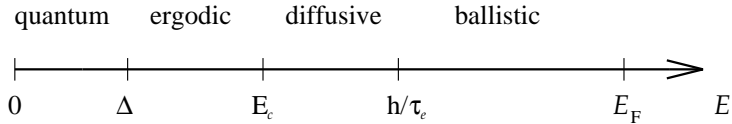}}
\caption{Relevant energy scales}
\label{f1:0}
\end{figure}
Although disordered systems have been mainly studied for their transport
properties, there has been in the recent years an increasing interest in
their spectral properties. This is in large part because of the development
of nanotechnologies which allows the fabrication of mesoscopic devices and
the detection of small signals. In particular, the discovery of persistent
currents
in mesocopic isolated rings has motivated the study of the statistical
properties of energy levels in the presence of an external parameter, a
Aharonov-Bohm (AB) magnetic flux, which breaks the Time-Reversal
 Symmetry (TRS)\cite{Levy90,Chandrasekhar91,Mailly93,Reulet95}.
This breakdown of TRS has the
effect of changing the spectral properties.
It has become also possible to measure the spectral correlations by
transport experiments in which it is possible to do a spectroscopy of the
levels\cite{Sivan94}.

Another motivation for the study of spectral correlations comes from an
argument due to Thouless which relates the
conductance to the typical variation of the energy levels in a change of the
boundary conditions\cite{Edwards72,Thouless74}. This argument links the
transport properties to the spectral properties and will be discussed in
details.
Among the signatures of the phase coherence on transport properties, a
very important one is the property of Universal Conductance Fluctuations.
In the mesoscopic regime , the conductance  is not a
self-averaging quantity and its distribution is characterized by a
 universal variance\cite{Stone85,Lee85,Altshuler85a,Altshuler86}
 \begin{equation}
 \langle  G^2 \rangle - \langle G \rangle^2 \sim
({e^2\over \hbar})^2\label{ucf}
\end{equation}
We shall see that this universal variance is closely related to the
{\it spectral rigidity}  mentioned above\cite{Altshuler86}.

 The
course is organized as follows:
the next section presents the microscopic model used for
disordered metals. It describes {\it non-interacting particles} moving in a
random potential.
It is shown that low energy spectral properties exhibit the universal
features of the Random Matrix  Theory (RMT) and that deviations exist above
the
energy scale $E_c$. In section \ref{sect3}, a semi-classical picture
relates
the spectral correlations to the description of a diffusive particle in a
random medium. Section \ref{sect4} is devoted to the persistent current in
a mesoscopic ring and its relation with the fluctuations of the local or
global density of states.
The curvature distribution and its relation with the conductance are
analyzed in section \ref{sect5}. Then it is seen that the parametric
correlations, namely the correlations between  energy levels at 
different values of an external parameter like an AB flux, exhibit
a universal behavior after rescaling of the energy and flux scales (section \ref{sect6}).

This course overviews only some aspects of the spectral fluctuations in
disordered metals, especially those using the semi-classical
approach\cite{Argaman93}. The subject has been covered by many aspects on recent
review papers\cite{Altshuler94,Kramer91,ALB91,Kramer93}. Throughout this
course,
unless otherwise specified, we shall set $\hbar=1,L^d =1$,   $\Delta
=1$ and we shall describe spinless electrons.

\section{Metal as a quantum chaotic system} \label{sect2}

The
study of spectral statistics in complicated systems has been initiated by
Wigner, Dyson, Mehta and others to describe the spectra of nuclei
\cite{Haake92,RMT,Mehta91,Bohigas86,Bohigas91}.
These authors describe the statistical properties of matrix Hamiltonians, with a
Gaussian distribution of the elements around a zero average.  A remarkable
feature of the Random Matrix Theory (RMT) is that the distribution of levels
depends only on the symmetry of the Hamiltonian. In particular, if the
Hamiltonian is invariant under time--reversal symmetry, the statistical ensemble
of matrices is invariant under orthogonal transformations and is called Gaussian
Orthogonal Ensemble (GOE). If the system is not invariant
 under time--reversal symmetry, for example in the presence of a magnetic
field, the statistical
ensemble  of matrices is invariant under unitary transformations and is
called Gaussian Unitary Ensemble (GUE).
 The RMT has
been applied to a variety of
very different physical situations in nuclear, atomic and molecular physics
\cite{Brody81,Haake92,RMT,Mehta91,Bohigas86,Bohigas91}.
The remarkable outcome  is that it gives a universal
description
of complex spectra and we shall use it to describe the spectra of
disordered  metals.

 The relevance of the RMT for a metal
 has been first pointed out by Gorkov and Eliashberg, in their study of
ac response of small metallic particles\cite{Gorkov65}. In
their work, the complexity  was not coming from the  many--body
nature of the Hamiltonian like in nuclear physics but from the
scattering of electrons
on the random shape of the boundaries, the system having the structure of a
billiard with rough boundaries. Here we describe a disordered metal where the electrons are
scattered elastically by fixed impurities.

Neglecting the electron-electron interactions, a  disordered
metal is usually described by a one particle  Schr\"{o}dinger
equation
 \begin{equation} {\cal H} \psi = {-\hbar^2 \over 2 m}
(\nabla  - i{e \vec A \over \hbar c})^2\psi + V({\bf r}) \psi = E \psi
\label{schrodinger} \end{equation}
$\vec A$ is the vector potential. The disorder potential is commonly
modeled as: \begin{equation}
\langle V({\bf r})\rangle = 0  \\
\langle V({\bf r}) V({\bf r}')\rangle  = {\delta({\bf r}- {\bf r'})\over 2
\pi \rho_0 \tau_e}
 \label{disorder}
\end{equation}
$\tau_e$ is the
elastic collision time.
This structure is convenient for analytical calculations and the
essential physics does not depend on the detailed choice of the disorder.
The discrete version of this Hamiltonian, called the Anderson tight-binding
Hamiltonian\cite{Anderson58}, is commonly used for numerical
calculations\begin{equation}
{\cal H} \psi_{i} = - \sum_{j} t_{ij} \exp(i {e \over\hbar c} \int_i^j A dl)
 \psi_j + V(i)  \psi_i
\label{tb} \end{equation}where the wave function
 $\psi_i$ is defined on the sites $i$ of a lattice.
In the simplest version of
this model, the $t_{ij}$ are taken as constant $t_{ij}=t$ and
$V(i)$ is a random variable with a uniform distribution in the
range  $[-W/2,W/2]$. This Hamiltonian,
first described  by Anderson,  exhibits localization
properties and a Metal-Insulator transition in 3D
\cite{Anderson58,Abrahams79}.    Fig. \ref{f2:spectra} shows a typical
spectrum in the diffusive and localized regimes, in the presence of an
external AB flux (see sect. \ref{sect4}). The goal of this
course is to describe the statistical properties of such spectra.
Let us first recall the usual quantities used to measure the level
fluctuations\cite{Mehta91,Bohigas86,Bohigas91} (The energies are measured
in units of the mean level spacing $\Delta$:
\begin{figure}[bh]
\begin{center}
{\parbox[t]{3.5cm}{\epsfxsize 3.5cm
\epsffile{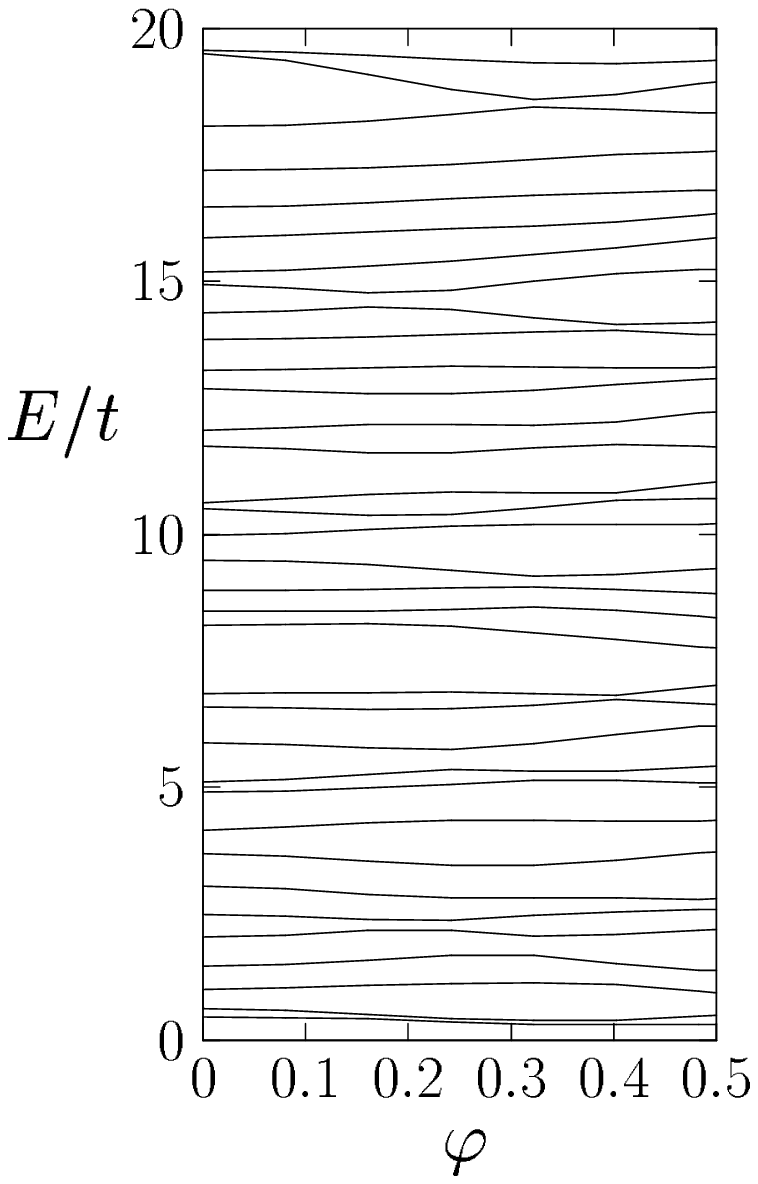}}
\hspace{-0cm} \parbox[t]{3.5cm}{\epsfxsize 3.5cm
\epsffile{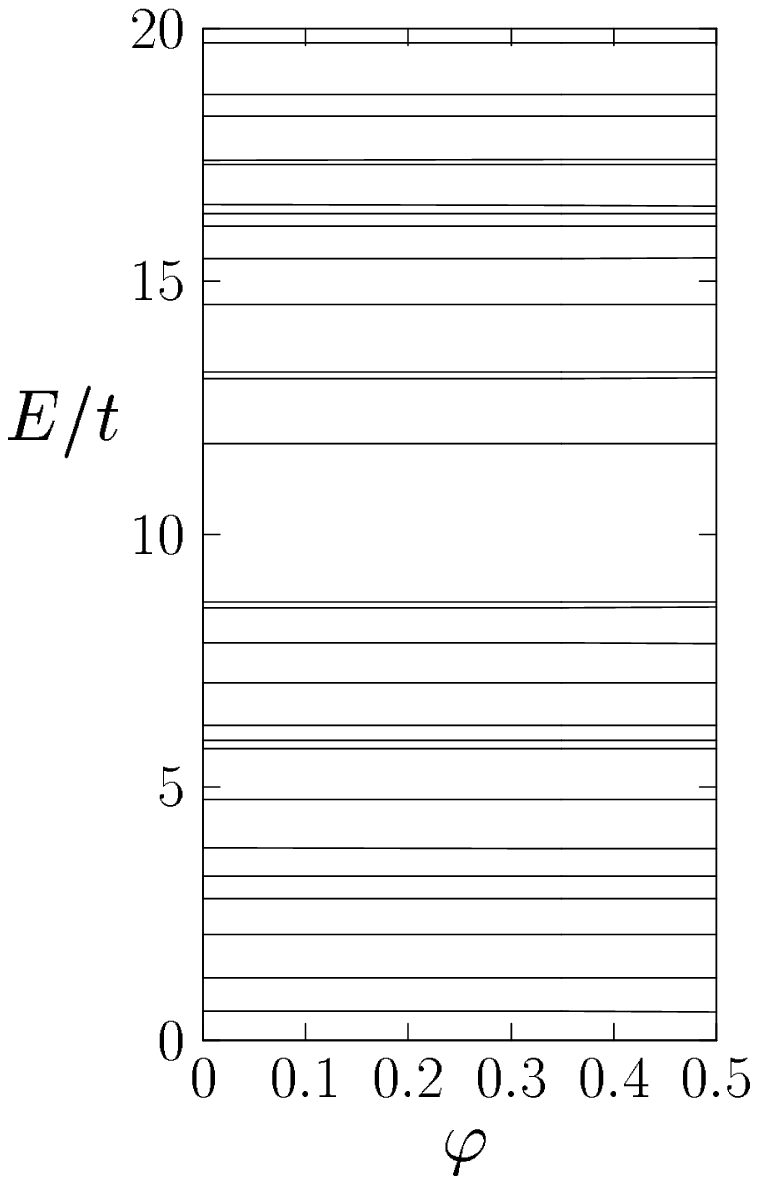}}
}
\end{center}
\caption{A typical spectrum of a metal in the diffusive regime (left),
and in the localized regime (right). $\varphi=\phi\phi_0$ is the
normalized Aharonov-Bohm flux, see sect. \protect\ref{sect4}.}
\label{f2:spectra} \end{figure}
 \begin{itemize}
\item  The distribution $P(s)$ of spacing $s$ between consecutive levels.
 In the RMT, it is well described
by the Wigner-surmise:
\begin{equation}
P(s) \propto s^\beta  \exp({-c_\beta s^2})
\label {wigner}
\end{equation}
where $\beta$ depends on the symmetry of the Hamiltonian. When
there is no correlation between levels, it has a Poisson behavior:
 \begin{equation}
P(s)= \exp(-  s)
\label {poisson}
\end{equation}
\begin{figure}[bth]
\begin{center}
{\parbox[t]{5cm}{\epsfxsize 5cm
\epsffile{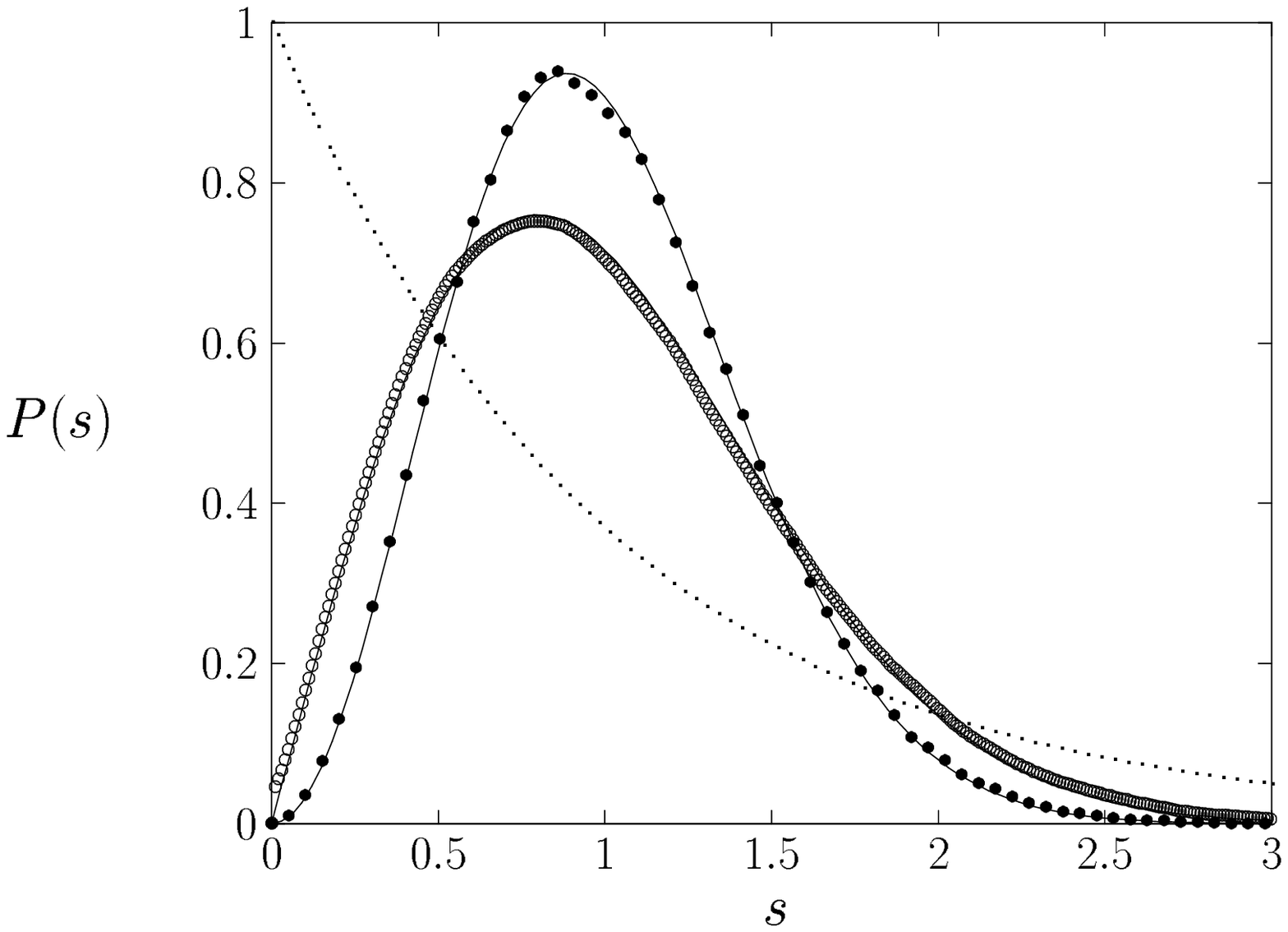}}
\parbox[t]{5cm}{\epsfxsize 5cm
\epsffile{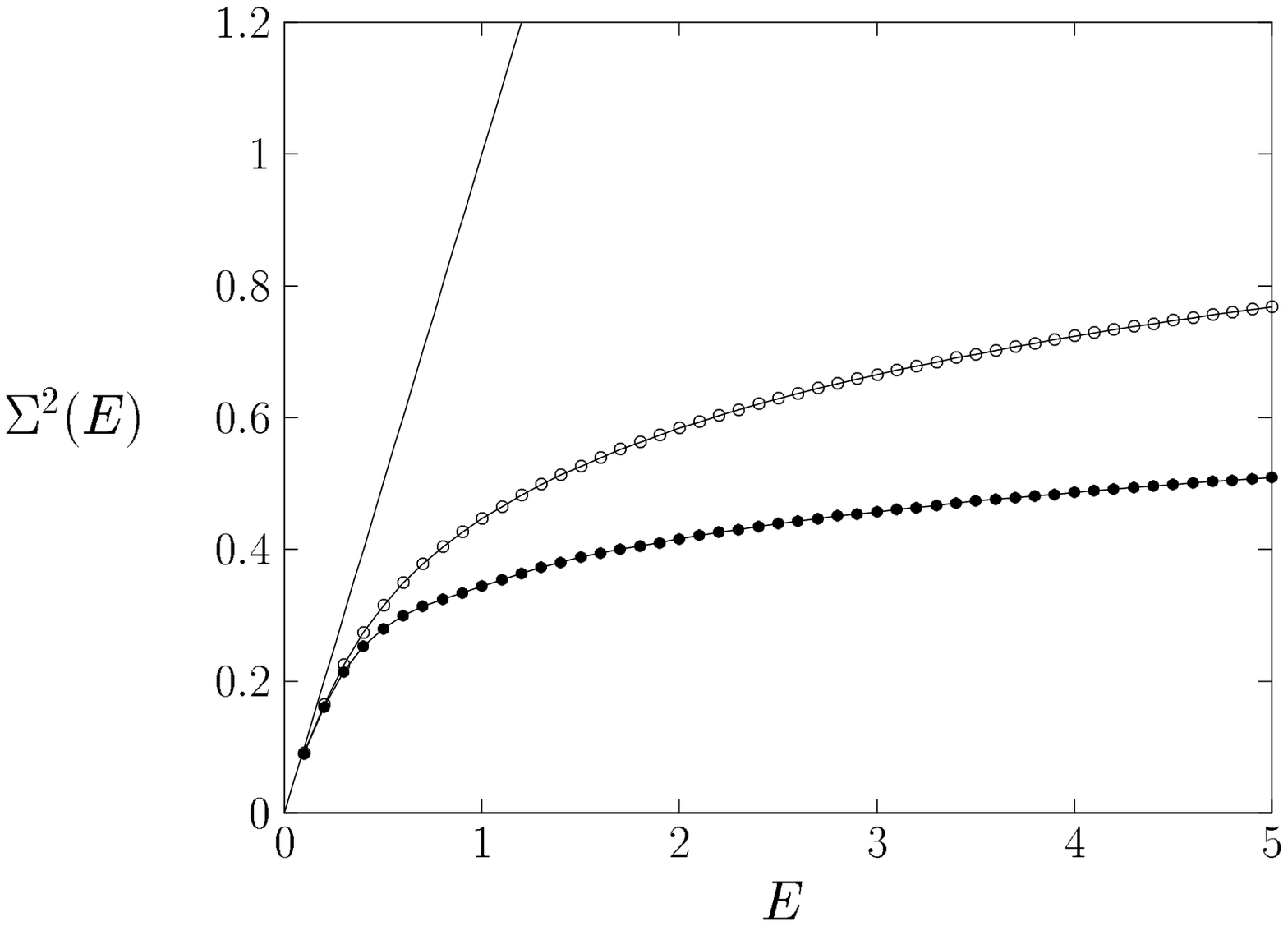}}
}
\end{center}
 \caption{$P(s)$ and  $\Sigma^2(E)$ for a metal in the diffusive regime,
with (black dots) and without (open dots) magnetic
flux\protect\cite{Dupuis91}. They are very
well described by the Random Matrix Theory. } \label{f2:rigidity}
\end{figure}
\begin{figure}[ht]
\begin{center}
{\parbox[t]{5cm}{\epsfxsize 5cm
\epsffile{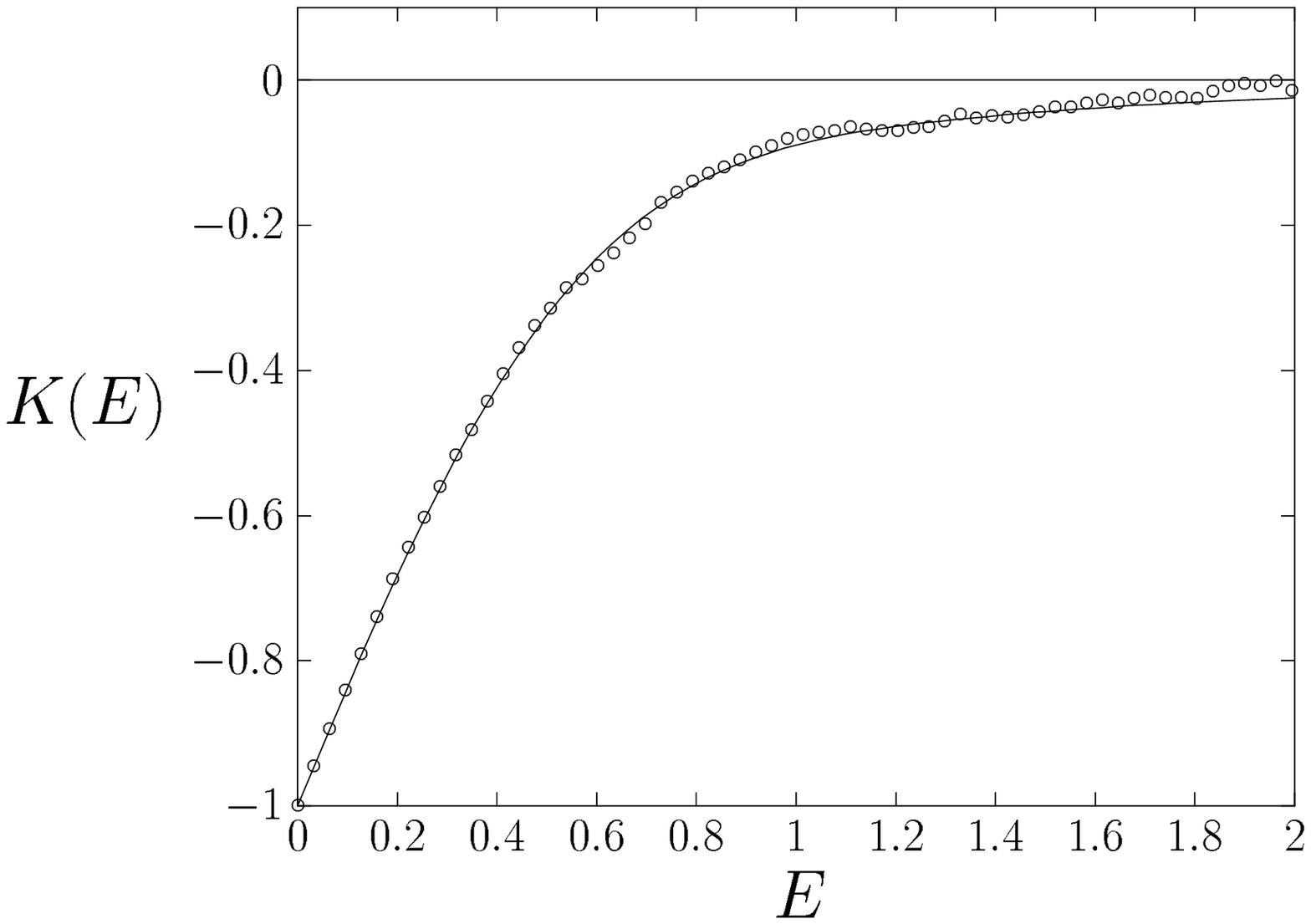}}   \hfill
 \parbox[t]{5cm}{\epsfxsize 5cm
\epsffile{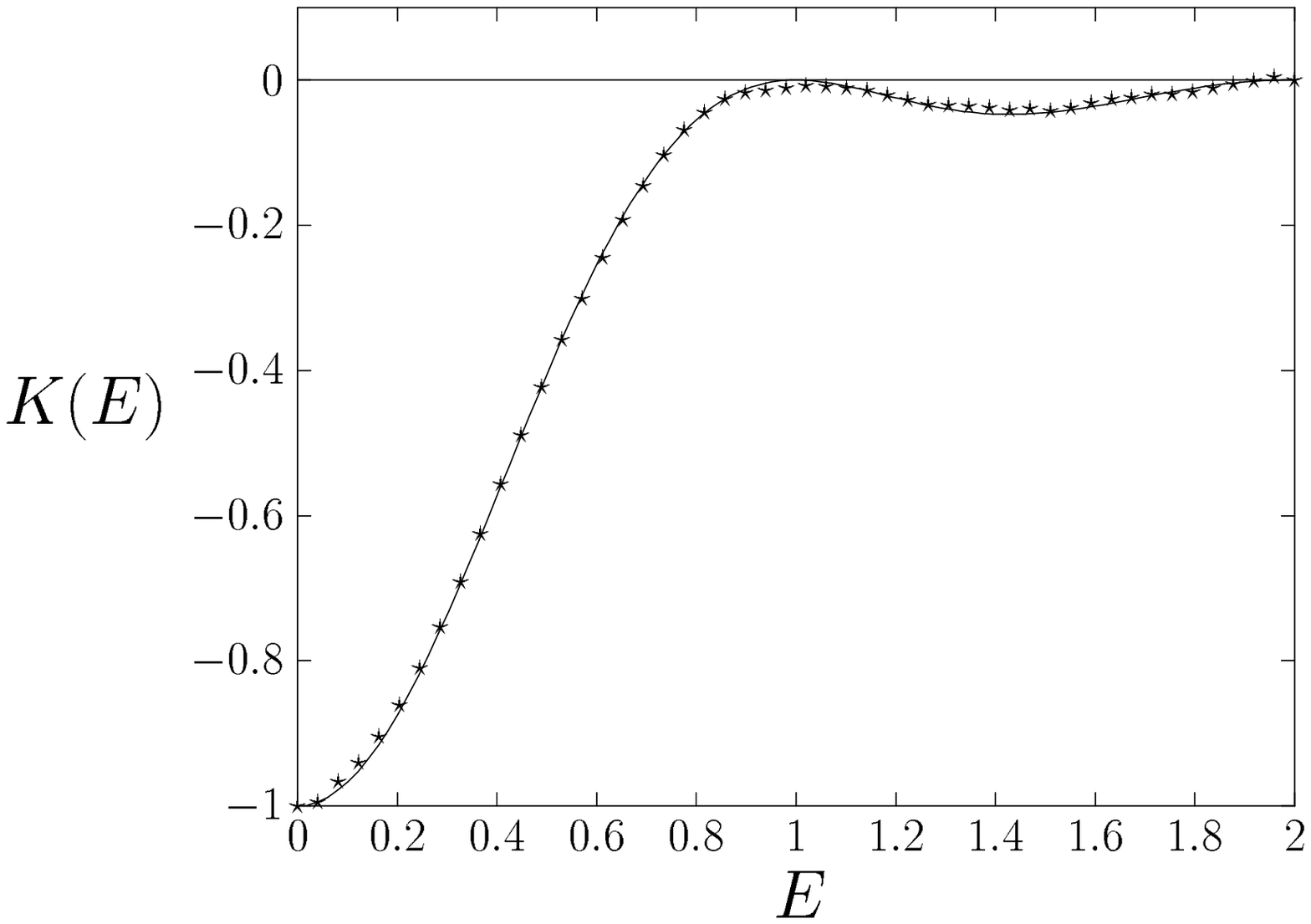}}
}
\end{center}
 \caption{Two-point correlation function  for a metal
in the diffusive regime, with (left) and without  (right) magnetic
flux\protect\cite{Braun95}. The continuous line is the RMT prediction
A $\delta$ function at the origin is not shown.} \label{f2:RdeE} \end{figure}
\begin{figure}[ht]
{\parbox[t]{5cm}{\epsfxsize 5cm
\epsffile{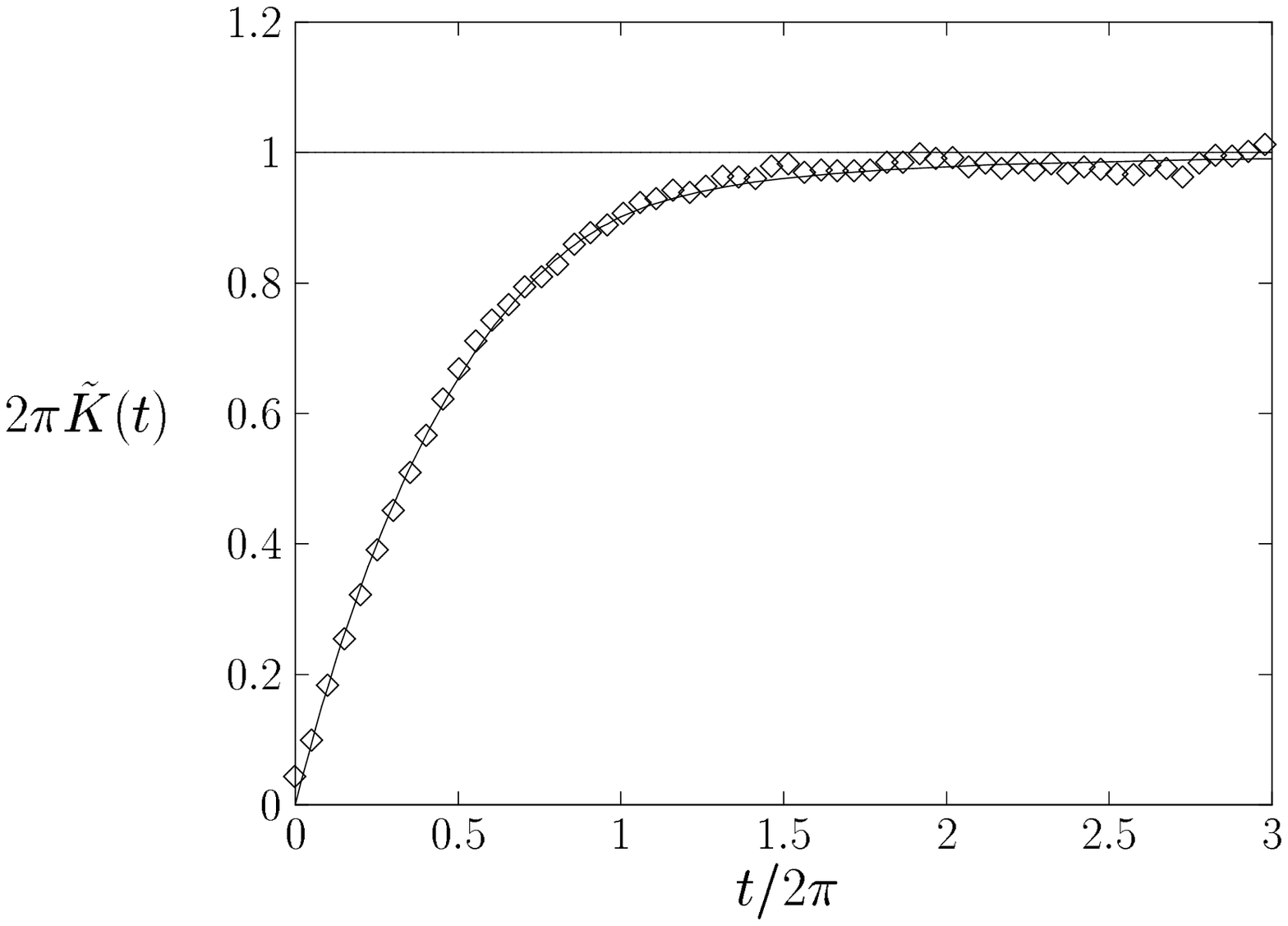}}
\hfill \parbox[t]{5cm}{\epsfxsize 5cm
\epsffile{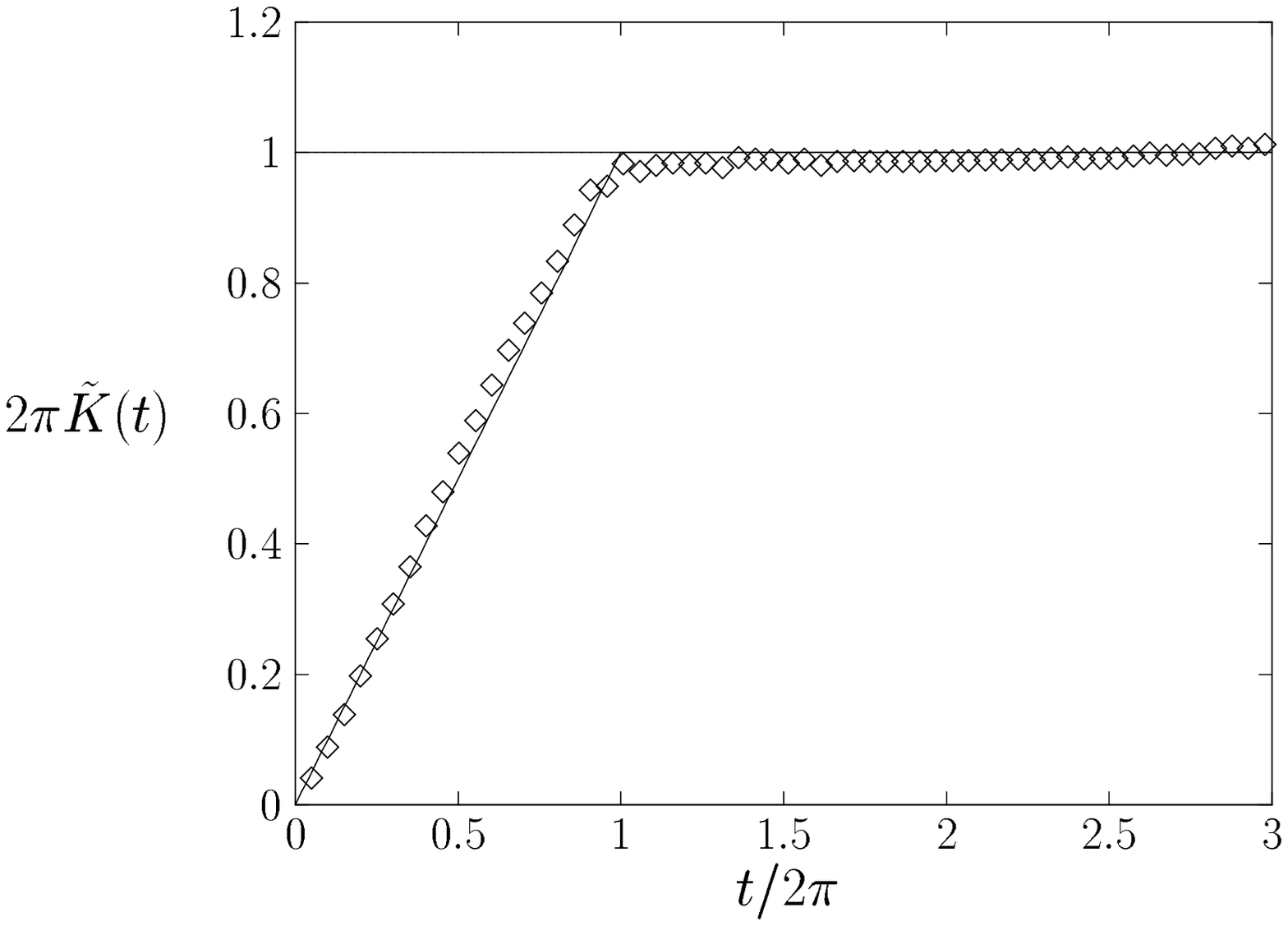}}
}
 \caption{The structure factor $\tilde{K}(t)$ for a metal in the diffusive
regime, with and without magnetic flux.} \label{f2:Kdet}
\end{figure}
\item  The two-point correlation function of the Density of States (DOS):
\begin{equation}
K(\ep_1,\ep_2)=
  \langle \rho(\ep_1)\rho(\ep_2) \rangle-  \rho_0^2
\label{KE}
\end{equation}
The average $\langle...\rangle$ is done on energy. Here the average will be
also
made on different disorder realizations. In the diffusive regime, these two
procedures are equivalent.
If, after averaging, the DOS is constant,  the function $K(\ep_1,
\ep_2)$ depends only on the
difference $\ep_1 - \ep_2$.   In the RMT and for large separation $\ep$, it
varies as $K(\ep) \rightarrow -1/(\beta  \pi^2 \ep^2)$. At $\ep = 0$, it has
a $\delta$ peak which describes the self-correlation of the levels. The
full expression is simple in the GUE case: $K(\ep) = \delta(\ep) -
\sin^2(\pi \ep)/(\pi \ep)^2$ (see fig. \ref{f2:RdeE}).
 \item The number variance
$\Sigma^2(E)$
 \begin{equation}  \label{variance}
\Sigma^2(E)\equiv \langle \delta N^2(E) \rangle = \langle N^2(E) \rangle -
\langle N(E) \rangle^2  \end{equation}
measures the fluctuation of the number of levels $N(E)$ in a strip
of width $E$.
By definition, the
number
variance can be   written in terms of this two-point correlation
function. 
\begin{equation}
\Sigma^2(E) = \int_0^E \int_0^E K(\ep_1 -\ep_2) d\ep_1 d\ep_2 =
 2 \int_0^E (E-\ep)K(\ep)
d\ep
\label{varK}
\end{equation}
In the RMT , for $E > \Delta$, it has a logarithmic behavior:
%
\begin{equation} \label{dn2RMT}
\Sigma^2(E)\simeq\frac{2}{\pi^2\beta}\ln(E/\Delta) + Cte
\end{equation}
Although the number variance is very frequently used in the literature, this
quantity is not always the most appropriate to describe the
correlation
because it is a double integral of the DOS-DOS correlation function
$K(\ep)$. Thus
the behavior of this quantity at an energy scale $E$ depends on the behavior
of the two-point correlation function for {\it all} energies smaller than
$E$.

\item The Fourier transform $\tilde{K}(t)$ is called the spectral form
factor
\begin{equation}
\tilde{K}(t) = {1 \over 2 \pi}\int K(\ep) \exp(i \ep t) d\ep
\end{equation}
The advantage of this
 quantity is to be directly related to
 the diffusive motion of a classical particle (sect. \ref{sect3}).
\end{itemize}
At small times $t \ll \tau_H$ where $\tau_H = 2 \pi
\hbar / \Delta$ is called the Heisenberg time, the form factor varies
linearly
with time $\tilde{K}(t) \rightarrow t /(2 \pi^2 \beta)$ and it saturates to a
constant value $1/2 \pi$ for $t \rightarrow \infty$. This constant is simply the
Fourier transform of the $\delta$ peak  which describes the
self-correlation of the levels\cite{Berry85}. 
Figures \ref{f2:rigidity},\ref{f2:RdeE},\ref{f2:Kdet} present numerical 
calculations of the different correlation functions, with the 
Anderson Hamiltonian\cite{Sivan87,Dupuis91,Braun95}. They are
very well fitted by the
 RMT.
Using the supersymmetric technique  to calculate $K(\ep)$ in
the microscopic
model (eq. \ref{schrodinger}), Efetov has  shown that its expression
actually coincides with the RMT result\cite{Efetov83}.
\begin{figure}[ht]
\epsfxsize 7cm
\centerline{
\epsffile{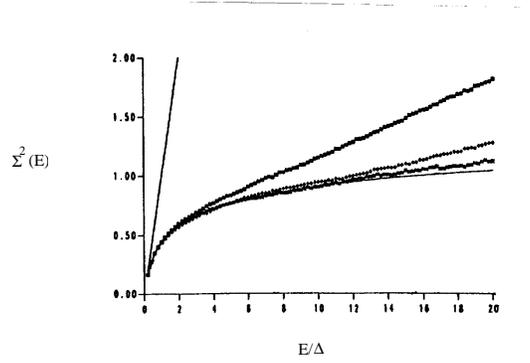}}
\caption{The  number variance exhibits deviations to RMT above $E_c$
} \label{f2:sigma2dev}
\end{figure}
\begin{figure}[bh]
\epsfxsize 7cm
\centerline{
\epsffile{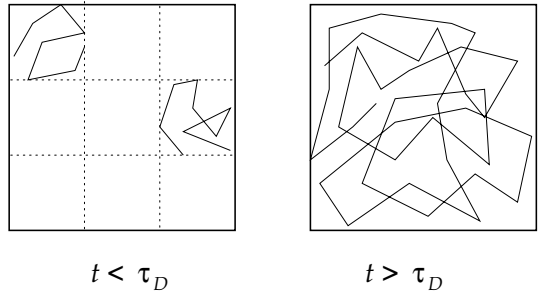}}
\caption{Schematic diffusion at small times ({\it diffusive regime})
and large times ({\it ergodic regime}) }
\label{f2:diffusion}
\end{figure}

However, fig. \ref{f2:sigma2dev} shows that above a given energy  which
 is size and disorder dependent and which has been
identified with $E_c$, the number variance exhibits
deviations and increases rapidly with energy.
These deviations from RMT are not surprising: the Anderson
Hamiltonian has a very different structure from a  random matrix.
It has many zero matrix elements. As we have seen in the introduction, there
are other energy scales not included in the RMT. Thus we
expect deviations from the universal behavior. RMT systems are
"ergodic" in the sense that their eigenfunctions
cover uniformly the phase space  and have no internal structure. This is
clearly not the case in a metal at small times  or large energies where a
diffusive particle cannot explore the entire space (fig.
\ref{f2:diffusion}).
More precisely, for small times $t_E \ll \tau_D$, i.e. large energies $E
\gg E_c$, the spatial
correlations extend on a scale $L_E = \sqrt{D t_E} = \sqrt{\hbar D
/E}$, so that the system consists of $(L/L_E)^{d}$ independent pieces, fig.
\ref{f2:diffusion}. Consequently, $\Sigma^2(E) \sim (L/L_E)^{d} \sim
(E/E_c)^{d/2}$ \cite{Altshuler86}.

In the next chapter, we
describe the specific properties of the metallic spectra and their deviations
from the universal
properties given by  the RMT.

 \section{Level correlations in a disordered metal}   \label{sect3}

We assume that the disorder is weak, so that the
 energy scales obey the hierarchy of eq. \ref{Escales}. Thus the
dimensionless conductance
$g \gg 1$. In this regime, we  want now to account for level
correlations within and beyond the RMT regime.

First it is  useful to introduce the following important operator
known  as the Green function or resolvent:
\begin{equation}
G^{R,A}(E) = {1 \over  E - {\cal H} \pm i 0}
\label{green1}
\end{equation}
In the site representation, the elements of the
Green function are given by
\begin{equation}
G^{R,A}({\bf r},{\bf r'},E) = \langle {\bf r} |
G^{R,A}(E)|{\bf r'}\rangle =    \sum_{n} {\phi^*_{n
}({\bf r}) \phi_{n}({\bf r'})\over E - E_{n} \pm i 0}
\label{green2} \end{equation}
$ E_{n}$ and $\phi_{n}$ are eigenenergies and
eigenstates of the Hamiltonian.
The Fourier transform $G^R({\bf r},{\bf r}',t)={1\over 2 i \pi} \int G^R({\bf r},{\bf r}',E) e^{-i E
t} d E$ has the following interesting property: Take a wave function
$\psi({\bf r},t=0)$,
solution of the time dependent Schr\"odinger equation (\ref{schrodinger}).
The time evolution of this wave function can be written as:
\begin{equation}
\psi({\bf r},t) = \int d {\bf r}' G^R({\bf r},{\bf r}',t) \psi({\bf r}',0)
\label{evolution}
\end{equation}
So $G^R({\bf r},{\bf r}',t)$ "propagates" a particle from $({\bf r}',0)$ to $({\bf r},t)$.
It results immediately that the probability  for a particle in
${\bf r}'$ at $t=0$, to be in ${\bf r}$ at $t$ is
\begin{equation}
 |G^R({\bf r},{\bf r}',t)|^2 =
G^R({\bf r},{\bf r}',t)  G^A({\bf r}',{\bf r},-t)
\end{equation}
If the particle is prepared at the energy $\ep_F$, the 
probability becomes\cite{Chakravarty86,Prigodin94}:
\begin{equation}
 P({\bf r},{\bf r}',t) ={1 \over 2 \pi}
 \int P({\bf r},{\bf r}',\omega) e^{- i \omega t} d \omega
\end{equation}with
\begin{equation}
 P({\bf r},{\bf r}',\omega) ={1 \over 2 \pi \rho_0}
 \langle G^R({\bf r},{\bf r}',\ep_F+{\omega \over 2}) 
 G^A({\bf r}',{\bf r},\ep_F-{\omega \over 2})\rangle
\end{equation}
 The DOS can be expressed
in term of the Green function as
\begin{equation}
\rho(E) = - {1 \over \pi }  \int
Im G^{R}({\bf r},{\bf r},E) d{\bf r}
\label{green3}
\end{equation}
It is also useful to define the {\it
local} DOS as
\begin{equation} \rho(r,E) = - {1 \over \pi }
Im G^{R}({\bf r},{\bf r},E)
\label{green4}
\end{equation}
The two-point correlation function can thus be
written as:
\begin{equation}
K(\ep_1,\ep_2) = {1 \over  2 \pi ^2 } \int \int d{\bf r} d{\bf r'}
 \langle G^{R}({\bf r},{\bf r},\ep_1)  
   G^{A}({\bf r'},{\bf r'},\ep_2)\rangle_c
\label{Ke1e2}
\end{equation}
where $\langle ...\rangle_c$ is the connected part of the average.
It is essential to notice that the Green function appearing in the DOS and
its correlation function is a {\it
diagonal} quantity (in space representation) which depends on the
propagation
of particles from some origin to itself.  All spectral quantities will thus
depend only on $G(\r,\r,E)$.

Another very important quantity which depends only on   the diagonal Green
function is the {\it return} probability to the origin $P(t)= P(\r,\r,t) $.
Its Fourier transform $P(\omega)$ is given by
\begin{equation}  \label{Pwreturn}
 P(\omega) ={V \over 2 \pi \rho_0}
 \langle G^R({\bf r},{\bf r},\ep_F+{\omega \over 2})
 G^A({\bf r},{\bf r},\ep_F-{\omega \over 2})\rangle
\end{equation}
Upon averaging, $P(\omega)$ is independent of the
position $\r$.

 One sees that  $K(\ep_1,\ep_2)$ and $P(\omega)$ have similar structures.
The goal of this section will be to relate these two quantities. To
calculate $P(t)$, we assume that the metal
is well described by a random potential in which the electrons experience a
diffusive motion of classical particles so that $P(\omega)$ is
a classical quantity equal to
$P_{cl}(\omega)$\cite{Chakravarty86,Bergmann84}. The 
probability to diffuse from $\r'$ to $\r$ is thus given by the 
solution of the classical diffusion equation i.e.
\begin{equation}\label{Pclt}
 P_{cl}(\r,\r',t)=  \sum_{\q} e^{-D q^2 t} e^{i \q(\r-\r')}
\end{equation}
where the diffusion modes $\q$ are quantized by the boundary conditions.
In the limit of an infinite system, it takes the familiar form
\begin{equation}
 P_{cl}(\r,\r',t)=
 {V \over (4\pi Dt)^{d/2} } e^{-|\r-\r'|^2/4Dt}
\end{equation}
where $D$ is the diffusion coefficient.
The return probability and its Fourier transform are thus given
by\cite{diffusion}:
 \begin{equation}  \label{cldiff}
 P_{cl}(t)=  \sum_{\q} e^{-D q^2 t} \ \ \ \ \ \
 P_{cl}(\omega)=  \sum_{\q} {1 \over -i \omega + D q^2}
\end{equation}

In  sect. \ref{Semi}, we shall show that
the two-point correlation function $K(\omega = \ep_1-\ep_2)$ is related
to the return probability $P(\omega)$.
\subsection{A brief reminder about weak-localization}
\label{WL}
Before
describing 
the structure of the spectral correlations,  let us first recall a brief
qualitative derivation of the first quantum correction to the classical
conductivity, called {\it weak-localization} correction
\cite{Chakravarty86,Bergmann84,Khmelnitskii84}. Linear
response theory shows that
 the d.c. $T=0K$ average conductivity\cite{Kubo57} has the
following structure:
\begin{equation}
\langle \sigma \rangle = - {e^2 \hbar^3 \over 2 \pi m^2 V } \int \int d{\bf
r} d{\bf r'}   \langle \partial_x G^{R}({\bf r},{\bf r'},\ep_F)
  \partial_{x'}  G^{A}({\bf r'},{\bf r},\ep_F)\rangle
\label{Kubo}
\end{equation}

By writing the Green function  as a sum of contributions from
classical
paths, as it will be explained in more details in the next
subsection (eq. \ref{Gscl}), the conductivity has the following structure:
\begin{equation}
\langle \sigma \rangle \propto \sum_{j,k} \langle B_j B^*_k e^{i (S_j -
S_k)/ \hbar}\rangle \label{WL1}
\end{equation}
where $B_j$  and $S_j$ are the amplitude and the action associated to each
path $j$. For most of the pairs of trajectories, $S_j - S_k > 2 \pi \hbar$,
so that their contribution to the conductivity cancels in average. 
The classical
conductivity is given by the sum of the intensities:
$\sigma_{cl} \propto \sum_{j} |B_j|^2$. However, there is a class of
trajectories which can also contribute to the conductivity: those which
form {\it closed loops}.
\begin{figure}[ht]
\epsfxsize 5cm
\centerline{
\epsffile{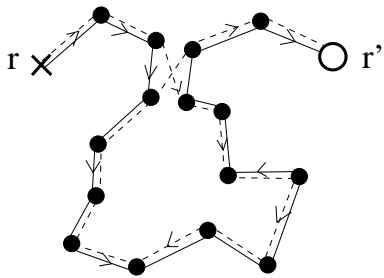}}
\caption{ The  two closed trajectories which make a closed loop are
time-reversal
symmetric. They give an additional contribution to the conductivity called
weak-localization correction. This correction is proportional to the
number of such closed trajectories.
}
\label{f3:weakloc}
\end{figure}
Such a loop can be traveled in clockwise or
 anti-clockwise directions. Both trajectories, $j$ and its time-reversed
$j^T$, have the same action, so that they interfere constructively.
As a result, in addition to the classical contribution, there is a
correction of the form  $\Delta \sigma \propto \sum'_{j} B_j B^*_{j^T}$
where the sum extends over the closed trajectories.
The sign of the correction is negative
 because the trajectories $j$ and $j^T$ have opposite
momenta. As a result there is an enhanced
probability to scatter in backwards direction.  The
conductivity is thus {\it reduced}. This effect is often
called {\it coherent backscattering or weak-localization}. This is a phase
coherent effect because
only trajectories of size smaller than $L_\varphi$ will contribute to
the correction.
The amplitude of the correction is proportional to the total number of
loops of any size. The number of loops of length $v_F t$ being proportional
 to the return probability
$P_{cl}(t)$, one deduces that the total correction is given by:
\begin{equation}
{\Delta \sigma \over \sigma} \propto - {\lambda_F^{d-1} v_F \over V}
\int_{\tau_e}^{\tau_\varphi} P_{cl}(t)  dt
\label{WL2}
\end{equation}
$\lambda_F^{d-1}$ is the transverse area associated to a semi-classical
trajectory. The integral is calculated between $\tau_e$, the smallest time
for diffusion, and $\tau_\varphi$, the time after which the electron looses
phase coherence. $\tau_\varphi$ is given by $L^2_\varphi = D \tau_\varphi$.
A magnetic field\cite{Altshuler80}
or a Aharonov-Bohm flux\cite{Altshuler81},  by breaking the time-reversal
symmetry, destroys the weak-localization correction\cite{Aronov87}.

In the next subsection, we calculate the spectral function $K(\omega)$, by
using very similar arguments.

 \subsection{Semi-classical description
of energy levels correlations}  \label{Semi}

We survey here a method which gives a  very physical description
of the level correlations because it directly  relates
them to the structure of the
classical motion of a diffusive particle.
More precisely, one can
 get a relation
between the spectral form factor $\tilde{K}(t)$ and the
classical return probability for a particle to return to the origin $P(t)$.
This description has been developed for diffusive
electrons by Argaman, Imry and Smilansky\cite{Argaman93} from
 ideas originally applied to other various classically chaotic
systems\cite{Dittrich90,Doron92,Smilansky92}.
We present here a brief and slightly different derivation.
The first step is to write the Green function as
a path integral
\cite{Feynman84,Feynman65}
\begin{equation} \label{gpi}
G^R({\bf r},{\bf r}',t)=\int_{{\bf r}'}^{{\bf r}} {\cal
D}[{\bf x}]\exp(\frac{i}{\hbar}S[{\bf x}])\,, \end{equation}
where $S[{\bf x}] = \int {\bf p} d {\bf x}$ is the classical action for the
path ${\bf x}(t)$ going from ${\bf x}(0)={\bf r}'$ to ${\bf x}(t)={\bf r}$.
The integral extends over all paths going from  ${\bf x}(0)={\bf r}'$ to
${\bf x}(t)={\bf r}$.
The classical paths given by  $\delta S[{\bf x}]/\delta {\bf x}=0$
give the main contribution to the integral
 (\ref{gpi}) so that the Green function can be written as
\begin{equation} \label{Gscl}
G^R({\bf r},{\bf r}',E)=\sum_j A_j({\bf r},{\bf r}',E)
 e^{i S_j({\bf r},{\bf r}',E)/ \hbar}\,, \end{equation}
This expression results from a stationary phase approximation near
classical paths\cite{Gutzwiller90,Gutzwiller91}. The amplitudes
$A_j$ result from a Gaussian
integration around the classical trajectories. Using the definition
(\ref{Pwreturn}) of the return probability $P(\omega)$, one gets
\begin{equation}
\label{Prwscl1} P(\omega) =   {1 \over 2 \pi \rho_0}\langle
\sum_{j,k} A_j({\bf r})A_k^*({\bf r})
e^{i [S_j(E+{\omega \over 2})-
S_k(E-{\omega \over 2})]/\hbar}\rangle\,, \end{equation}
This sum is done over {\it closed} paths. We have defined
$A_j({\bf
r}) \equiv A_j({\bf r},\r)$. Moreover, for a closed path, the
action is independent of the starting point.
The main energy dependence is contained in the phase factors and
we have suppressed it in the amplitudes $A_j$.
It has been argued by  Berry that the phase factors with large actions such
that $S_j - S_k \gg 2 \pi \hbar$ should cancel on
average\cite{Berry85,Berry91}.
As a result, in the absence of time reversal symmetry, only the diagonal
terms are kept in the sum. Using the relation
$T_j = d S_j/ d E$ between the period and the energy dependence of the
action for a closed path, the phase factors can be expanded as
$S_j(E+\omega/2) = S_j(E) + \omega T_j /2$, so that one obtains:
\begin{equation} \label{Prwscl2}
P(\omega) = {1 \over 2 \pi \rho_0}\langle
\sum_{j} |A_j({\bf r})|^2
e^{i \omega T_j}\rangle  \end{equation}
By Fourier transformation, the return probability can be written as:
\begin{equation}   \label{Prtscl}
P(t) = {1 \over 2 \pi \rho_0} \langle
\sum_j |A_j({\bf r})|^2
\delta(t -  T_j) \rangle     \\
\end{equation}

After disorder averaging, this quantity is independent of the origin
$\r$.
We now turn to the calculation of the two-point correlation function, that
we write in a symmetrized form. Using the expression in terms of the Green
functions, doing the same manipulations as for the return probability,
one obtains
\begin{eqnarray} \label{Ktscl}
\tilde{K}(E,t) &=& {1 \over 2 \pi} \int \langle \rho(E+{\ep \over
2})\rho(E-{\ep \over
2})\rangle e^{i \ep t}  d\ep          \nonumber \\
 &=& {1 \over 2 \pi^2}
\langle\sum_{j}
\int \int d \r d\r'A_j(\r)A_j^*(\r') \delta(t -  T_j) \rangle    \nonumber
\\ &=& {1 \over 2 \pi^2} \langle\sum_{j}
|A_j|^2 \delta(t -  T_j) \rangle \,, \end{eqnarray}
where $A_j \equiv \int  A_j(\r) d \r$. Clearly, the
expressions for $P(t)$
and $\tilde{K}(t)$ are very similar. To relate them, it is now important to
make the following
distinction. The semi-classical motion is characterized by an ensemble of
closed {\it trajectories}. $P_j = |A_j({\bf r})|^2$ is the probability to be
on the trajectory $j$ with origin  $\r$.  Many other
trajectories
follow the same path but they start from different points $\r$ and $\r'$.
This ensemble of trajectories, who have the same action, is called an
{\it orbit}.
\begin{figure}[ht]
\epsfxsize 10cm
\centerline{
\epsffile{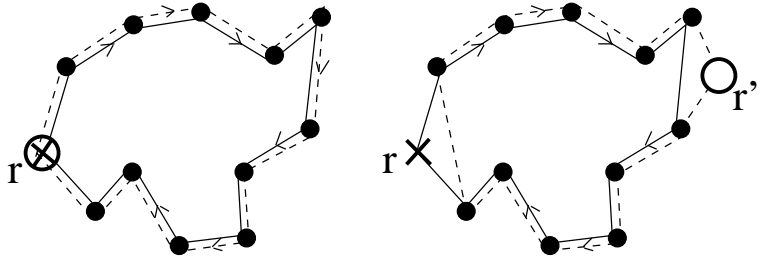}}
\caption{a) Two identical trajectories starting from the same origin
and belonging to the same orbit: these trajectories contribute to $P(t)$.
b) Two trajectories with different origin belonging to the same orbit:
these trajectories contribute to $\tilde K(t)$.} \label{f3:orbits}
\end{figure}
The probability to be on a given orbit $j$ is
$|A_j|^2 = \int \int d\r d\r' A_j(\r) A^*_j(\r')$.
One sees that for one given $\r$, the
possible choice for $\r'$ on the orbit $j$ is proportional to the
volume
of this orbit.
 This "semi-classical"
volume is $\lambda_F^{d-1} v_F T_j$ were $\lambda_F^{d-1}$ is the section
of the classical orbit and $v_F T_j$ is its length.
Recalling that  $\lambda_F^{d-1} v_F$  is proportional to the inverse DOS,
one deduces (this simple argument cannot reproduce here the correct
prefactor):
\begin{equation} \label{Ozorio} |A_j|^2 =  {1 \over 4 \pi \rho_0} P_j T_j
\end{equation}
This result  has been obtained for the
first time by Hannay and Ozorio de Almeida in the ergodic case ($P_j =1$)
\cite{Hannay88,Berry91}.
After summation over the orbits, comparing eqs. \ref{Prtscl} and
\ref{Ktscl}, one obtains:
\begin{equation}   \label{tPt}
\tilde{K}(t)  =  {1 \over 4 \pi^2 } t P_{cl}(t)
\end{equation}

This semi-classical
expression is believed to be valid for small
orbits , with period $t$ smaller than the quantum (or Heisenberg) time
$\tau_H = 2 \pi \hbar / \Delta$\cite{Argaman93,Berry91}.
Fourier transform of eq. \ref{tPt}
gives
\begin{equation}   \label{wPw}
 K(\omega)  =  {1 \over 2 \pi^2 } \Im {\partial P_{cl}(\omega) \over
 \partial \omega}=
-{1 \over 2 \pi^2} \Re \sum_q {1 \over (-i \omega + D q^2)^2}
\end{equation}
and the number variance is obtained from double integration
(eq. \ref{variance}):
 \begin{equation} \label{sigma2pt}
\Sigma^2(E)  =  8 \int_0^\infty dt  {\tilde{K}(t)  \over  t^2
}\sin^2({E t \over 2})=  {2  \over  \pi^2} \int_0^\infty dt
{P(t) \over t }\sin^2({E t\over 2})
  \end{equation}

In the above diagonal approximation, eq. \ref{Prwscl2}, we kept the terms
where $k=j$. However, in the case where the system  is  time-reversal
invariant, for each orbit $j$ there is an orbit which is
reversed by time inversion   $k=j^T$ and which have the same 
action. Thus, the non-diagonal terms $k=j^T$ also survive  on
average. This {\it interference} term is not taken into account 
in the classical diffusion so that the semi-classical 
probability to return to the origin is actually {\it twice} the 
classical probability $P(t) =2 P_{cl}(t)$. This description is 
called {\it semi-classical} because it is essentially a classical
picture in which the quantum mechanics only appears through a 
phase coherent classical contribution. The two 
contributions are called in the diagrammatic language the 
diffuson (diagonal) and the cooperon (interference)   
contributions. Eq. \ref{wPw} has been obtained for
the first time directly by Altshuler and Shklovskii, using a 
diagrammatic calculation\cite{Altshuler86}.

In order to take into account the loss of coherence due to dephasing
events, an exponential damping $e^{-\gamma t}$ has to be added in the
interference term of $P(t)$ which cuts the contribution of long orbits, so
that the diffusion pole
in \ref{wPw} has a gap $-i \omega + D q^2 \rightarrow -i \omega + D q^2
+ \gamma$. The inverse scattering time $\gamma$ is related to the
 coherence length $L_\varphi$: $\gamma = \hbar D / L_\varphi^2$
Moreover, the semi-classical approximation breaks down for small
energy scales $\omega < \Delta$ so that, even without damping,  the
divergence in the diffusion pole has to be regularized by
the transformation $ \omega  \rightarrow  \omega +   i
\gamma_\Delta$ where $\gamma_\Delta$ is an energy scale of the order
of $\Delta$\cite{Dupuis91}.
The correlation function exhibits clearly two distinct
regimes:

When $\omega <E_c$ or $\tau > \tau_D$, the diffusion is uniform in the
sample (fig. \ref{f2:diffusion}): $P(t)=2/\beta$ so
 that $\tilde{K}(t)$ varies linearly in time
$\tilde{K}(t) =  t /(2 \beta \pi^2)$. By Fourier transform, one has
$K(\omega)=-1/
\beta \pi^2 (\omega+i\gamma_\Delta)^2$ and the  number variance varies
logarithmically:
$\Sigma^2 = (2 / \beta \pi^2) \ln (E / \gamma_\Delta)$. This is the {\it
ergodic}
regime well described by the RMT. It is obtained by taking only the $\q=0$ 
contribution in the sum \ref{wPw}. This contribution is called 
the {\it uniform} or {\it zero mode}. It should be emphasized that this
semi-classical approximation cannot describe properly time scales
close to $\tau_H$ or energy scales $\omega \leq \Delta$.

Keeping only the
diagonal terms corresponds to the GUE
case ($\beta=2$) and the interference terms double the contribution to the
correlation function ($\beta=1$). This explains why the fluctuations in the 
GOE case are about twice as large as in the
unitary case.

In the opposite limit (fig. \ref{f2:diffusion}), when
$\omega \gg E_c$ or $\tau \ll \tau_D$, the return probability
 depends on the space dimensionality $d$ :  $P(t) \propto V /(D t )^{d/2}$
so that $\tilde{K}(t) \propto   t^{1-d/2}/D^{d/2}$ and, for $\omega \gg
E_c$:
 \begin{equation} \label{Kas}
K(\omega) \propto   -  {1\over \beta \omega^2} \ \left( {\omega \over
g}\right)^{d/2}  \cos({\pi
d \over 4})
\end{equation}
This leads to a power law dependence of the
number variance\cite{Altshuler86}
  \begin{equation}  \label{as}
  \Sigma^2(E) \propto {1 \over \beta} \left({E \over E_c}\right)^{d/2}
\end{equation}
It is remarkable that the sign of the correlation function is now dimension
dependent. In three dimensions, $K(\omega)$ varies at
large energies  positively like $+1 / \sqrt{\omega}$
instead of $-1/\omega^2$ for RMT, meaning that the levels tend to
{\it attract}  each other at
large energies\cite{Jalabert93}.
 The power law regime of the variance has been found numerically recently.
However, the deviation of the correlation function from RMT is quite hard to
observe numerically in the metallic regime\cite{Kravtsov94b,Andreev95,Braun95}
when $g \gg 1$.

Finally, when $E > \hbar /\tau_e$, i.e. $t<\tau_e$, the motion is not
diffusive anymore and becomes ballistic. In this regime, the spectral
rigidity is very weakly dependent on $E$\cite{Altland93a}.

\subsection{Breakdown of time-reversal symmetry, Aharonov-Bohm flux}
\begin{figure}[ht]
\centerline{
\epsfxsize 5cm
\epsffile{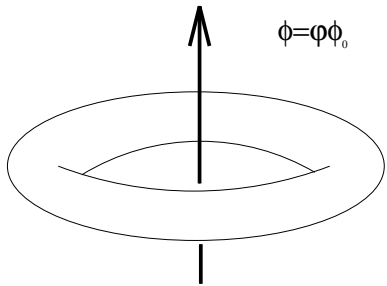}}
\caption{A ring pierced by a Aharonov-Bohm flux}
\label{f3:ring}
\end{figure}

The above section has stressed the importance of the 
interference effects in the origin of spectral rigidity. They 
are affected by the presence of an external parameter which 
breaks the time-reversal symmetry, for example a
Aharonov-Bohm (AB) magnetic flux: one considers a disordered metal having
the geometry of a quasi-1D ring of perimeter $L$ pierced by an
AB flux $\phi$,
fig. \ref{f3:ring}. It is  assumed that the
vector potential $A$ is a constant $\phi/L$ in the ring. The
energy spectrum is solution of the Schr\"odinger equation
(\ref{schrodinger}) with the periodic boundary condition
$\psi(x+L) = \psi(x)$ where $x$ is the coordinate along the ring.
A gauge
transformation    $\psi'(x) = \psi(x) e^{i e \int A dl / 
\hbar}$ removes the flux from the Hamiltonian: ${\cal H}(\phi) 
\psi = {\cal H}(0) 
\psi'$ but the new wave function $\psi'$ obeys a new boundary
condition  $\psi'(x+L)=\psi'(x)
e^{2i\pi \phi/\phi_0}$,
 where $\phi_0=h/e$
is the flux quantum.
{\it An external Aharonov-Bohm flux is thus mathematically
equivalent to a change in the boundary condition}\cite{Byers61}.
Moreover, it breaks the time reversal symmetry
and must lead to a change in the spectral statistics, from 
 the 
orthogonal to the unitary symmetry.

The transition between the two symmetries
 for random matrices has been
 considered by
Pandey and Mehta \cite{Pandey83,French88}.
They have used an
interpolating
 ensemble   of matrices
of the form $H= H(S)
 + i\alpha H(A)$
where $H(S)$ and $H(A)$ are real symmetric and
antisymmetric matrices of dimension $N$ and variance $v^2$. ($\alpha=0$) is
the orthogonal case and ($\alpha=1$) describes the unitary symmetry.
These authors have shown that
the transition between
 GOE and GUE
is
 driven by the
single parameter $\Lambda = v^2  \alpha^2 / \Delta^2$.
For Gaussian matrices, the average interlevel spacing is not a constant
in the spectrum so that the transition depends on the position
in the spectrum\cite{Pandey83,French88}.
For a Gaussian matrix of size $N$ and variance $v^2$, the mean
level spacing  in the band center 
is given by $\Delta= \pi v / \sqrt N $ so that
the parameter which drives the transition is the combination
$\Lambda = N \alpha ^2/\pi$. For example, for small times $t \ll \tau_H$,
Pandey and Mehta found that the form factor is given by
 \begin{equation}\label{PM}
\tilde{K}(t,\alpha) =  {t \over  4
\pi^2 }[1 +
e^{- 4 \pi \Lambda \Delta t}]
 \end{equation}
and interpolates between GOE ($\Lambda=0$) and GUE ($\Lambda \sim 1$). To
relate these parameters to
those of our physical problem, let us return to the semi-classical
description. With the above
gauge transformation,
the diagonal Green function gets an additional phase factor:
\begin{equation} \label{Gsclphi}
G^R({\bf r},E,\varphi)=\sum_j A_j({\bf r})
 e^{i [ S_j/\hbar + 2 \pi m_j \varphi]}
\end{equation}
where $\varphi=\phi/\phi_0$ and $m_j$ is the winding number of the
trajectory $j$. Following the same steps as in section \ref{Semi}, one finds
 \begin{equation}
\label{Prwsclphi1} P(\omega,\varphi) =  {1 \over 2 \pi \rho_0}\langle
\sum_{j} |A_j({\bf r})|^2
e^{i\omega T_j} [1 +
e^{i 4 \pi m_j \varphi}] \rangle
\end{equation}
and its Fourier transform:
\begin{equation}
P(t,\varphi)=  {1 \over 2 \pi \rho_0}\langle
\sum_{j} |A_j({\bf r})|^2
\delta(t -  T_j) [1 + \cos(4 \pi m_j \varphi)]\rangle
\label{Prtsclphi}  \end{equation}

The diagonal terms $k=j$ are unchanged and the terms $k=j^T$ get
a phase $4\pi m_j \varphi$, since, after one turn around the loop,
one trajectory picks up a phase $2\pi  \varphi$ and its time reversed
picks up a phase $-2\pi \varphi$. Classifying the trajectories with
respect to their winding number, one finds:
\begin{eqnarray}\label{Ptphi}
P(t,\varphi)&=& P_{cl}(t,0)+P_{int}(t,\varphi)    \nonumber \\
&=&  {L \over  \sqrt{4  \pi D t}}
\sum_{m=-\infty}^{\infty} e^{- m^2 L^2 / 4 D t} [1 + \cos(4 \pi m \varphi)]
\label{Prtsclphi2}  \end{eqnarray}

\begin{figure}[ht] \centerline{
\epsfxsize 6cm
\epsffile{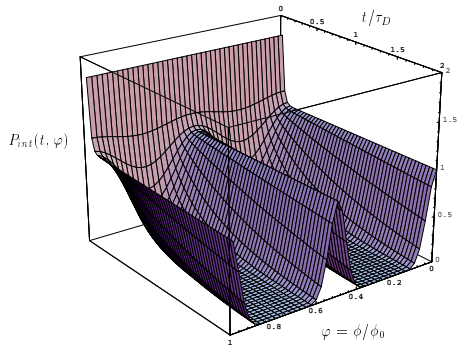}}
\caption{Interference part of the return probability
} \label{f3:pdet}
\end{figure}

The first term is the classical return probability and the second is
the interference term which oscillates with period $\phi_0/2$.
We have assumed that along the transverse directions, the return probability is
time independent so that our expression is one-dimensional.
The flux dependent part of $P(t,\varphi)$ is shown on figure \ref{f3:pdet}.
This figure clearly
exhibits the two regimes,   $t < \tau_D$ where the motion is diffusive and
the flux dependence is small, and $t > \tau_D$ where the motion is ergodic
and the  flux dependence is large.

Let us first describe the limit  where  $t \gg \tau_D$ and
$\varphi \ll 1$ so that the sum over winding numbers can be
replaced by an integral. This is the
{\it zero-mode approximation} which corresponds to the {\it ergodic regime}.
The return probability becomes:
\begin{equation}\label{Prtsclflux}  
\tilde{K}(t,\varphi) = {1 \over 4 \pi^2} t P(t,\varphi)=  {t \over  4 \pi^2
}[1 + 
e^{- 16 \pi^2 E_c \varphi^2 t}] 
\end{equation}
This result is similar to the one found in the RMT, with the mapping:
\begin{equation}\label{mapping}
 \Lambda ={N \alpha^2 \over \pi} = 4 \pi {E_c\over \Delta} \varphi^2
 \end{equation}
 This mapping is quite natural: $E_c / \Delta$,  which gives the
range of the RMT correlation, is related to the dimension $N$ of the random
matrices and $\varphi$ plays the role of the symmetry-breaking parameter
$\alpha$\cite{Dupuis91}.

We can then use the exact RMT result
to calculate any correlation function  in a flux for any energy 
range $\ep < E_c$. All the  correlation functions  are thus  universal
functions of the combination of parameters $E_c
\varphi^2$\cite{Dupuis91,Akkermans92b}.
\begin{figure} \label{f3:crossover} 
{\parbox[t]{5.6cm}{\epsfxsize 5.6cm
\epsffile{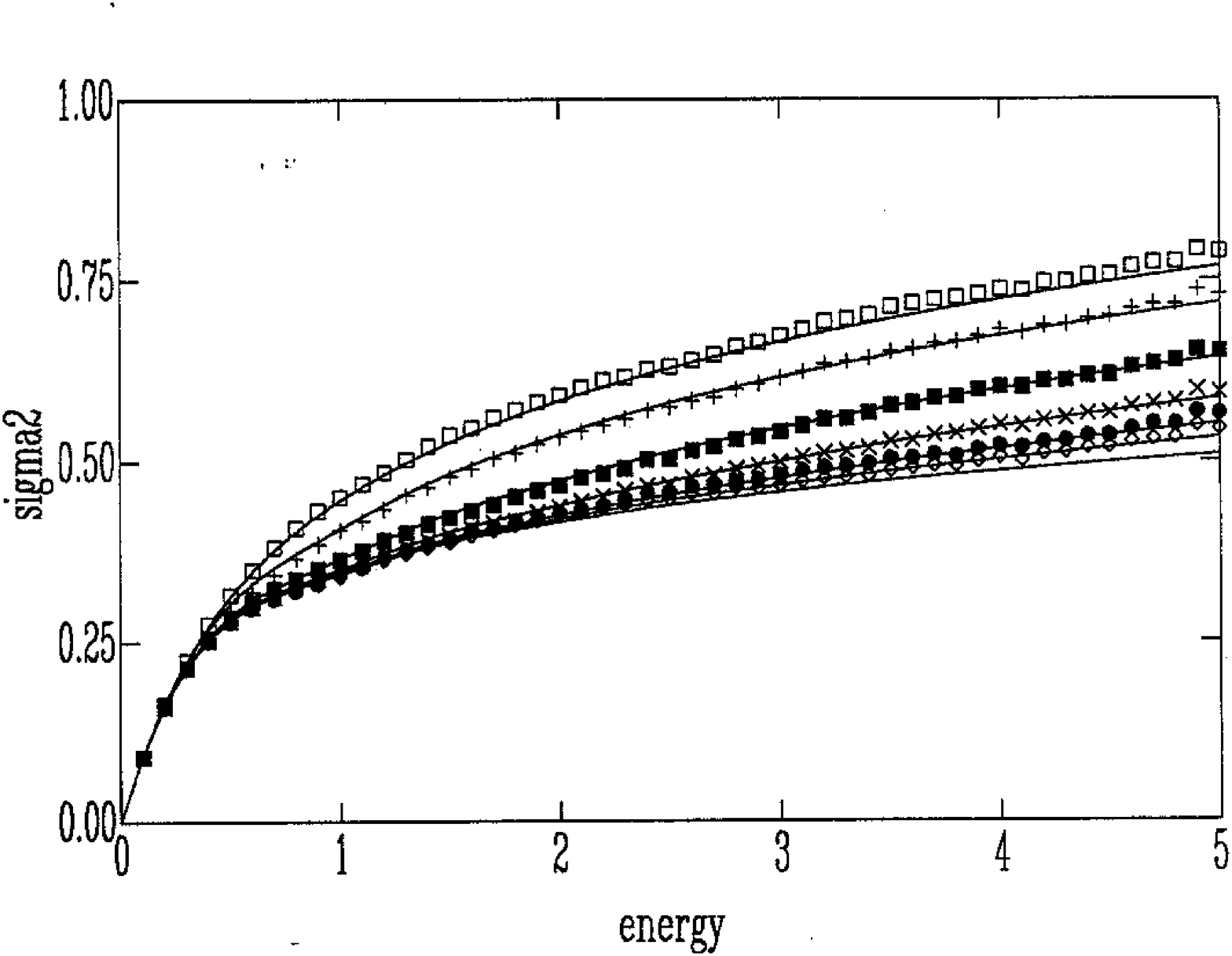}}
\hfill \parbox[t]{5.6cm}{\epsfxsize 5.6cm
\epsffile{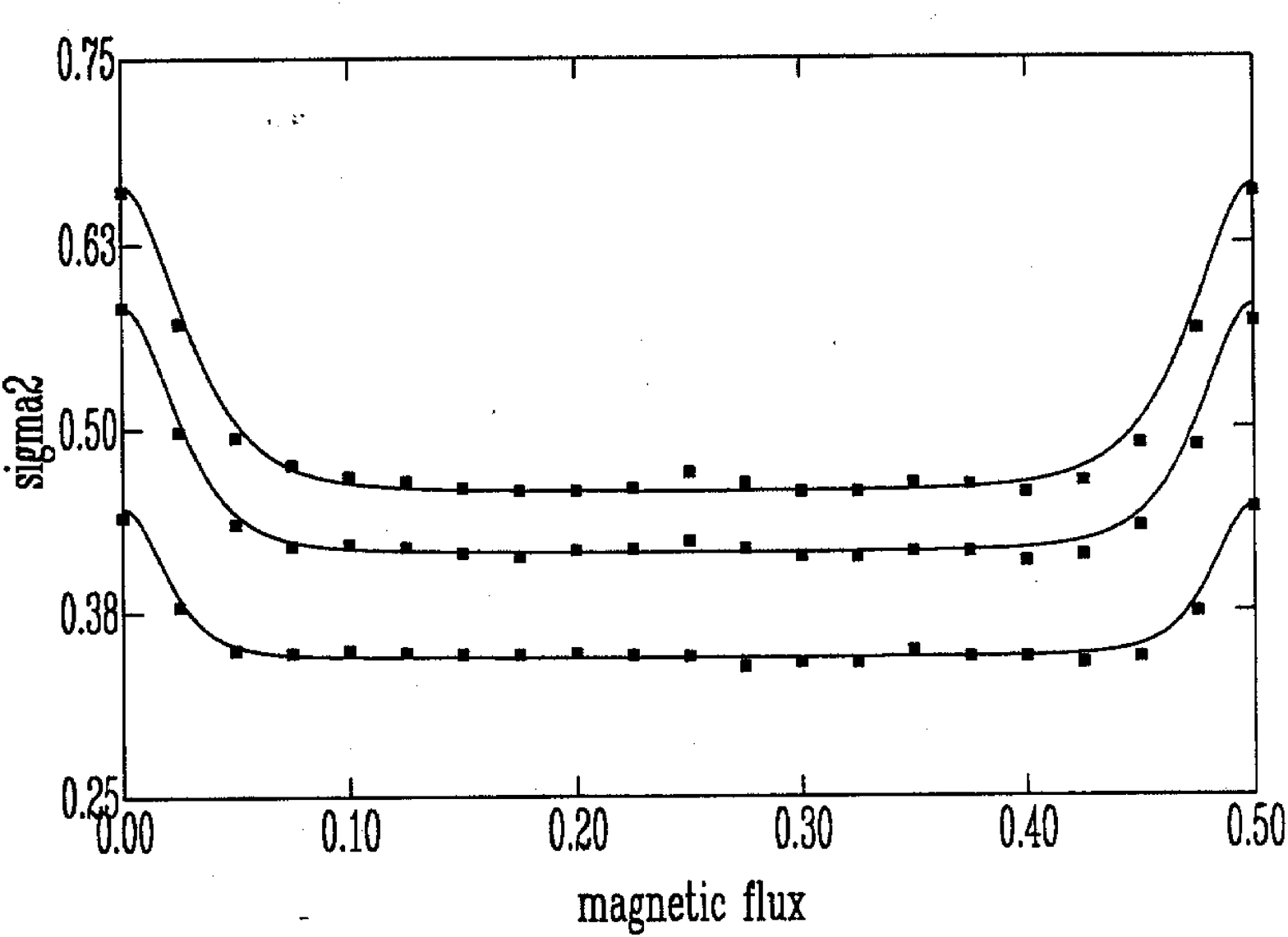}}}
\caption{Variance number $\Sigma^2(E,\varphi)$ for different values of
the flux $\varphi$ and comparison with RMT ({\it left}).
$\Sigma^2(E,\varphi)$
versus flux for a given energy ({\it right}): note that this variation is
reminiscent of the variation of $P(t,\varphi)$ vs $\varphi$
when $t \gg \tau_D$, fig. \protect\ref{f3:pdet}} \end{figure}
Using a supersymmetric approach, Altland {\it  et al.} have indeed
directly found
that the two point correlation in the cross-over regime is well
described by the RMT correlation function\cite{Altland93b}.

This dependence (\ref{Prtsclflux}) of the form factor with the flux has a
very simple origin: diffusive pairs of
orbits accumulate a typical phase $\langle \Phi^2 \rangle = \langle (4 \pi m
\varphi)^2 \rangle$ where $\langle m^2 \rangle$ is the typical winding
number. In the diffusive regime,  $\langle m^2  \rangle = 2 D t /L^2 = 2 E_c
t$. $\langle \cos\Phi \rangle = e^{-\langle \Phi^2 \rangle /2}=
e^{- 16 \pi^2 E_c \varphi^2 t}$.
The symmetry breaking depends on the length of the trajectories. For a
trajectory of time $t$, TRS is broken for $\Phi \sim 2 \pi$, i.e.
$\varphi_c \sim 1/\sqrt{E_c t}$. The characteristic
flux associated with an energy $\ep$ is thus $\phi_c \sim \sqrt{\ep/E_c}$.

It should be emphasized that this argument holds even for 
the ballistic regime in the non-integrable case, for example
for billiards in an AB flux. In that case, the motion is 
ballistic but the accumulation of the phase with time is still 
diffusive\cite{Berry77,Serota94,Bohigas95,Altland95}.

Using the relation \ref{variance}, we obtain the number variance:
\begin{equation}
\Sigma^2(E) = {1 \over 2 \pi^2} \ln(1+{E^2 \over \gamma_\Delta^2})+
{1 \over 2 \pi^2}
\ln\left(1+{E^2 \over (\gamma_\Delta+ 16 \pi^2 E_c \varphi^2)^2} \right)
\end{equation}
This fits perfectly the variation of the number variance with the
flux (fig. \ref{f3:crossover}). For $E_c \varphi^2 \gg \Delta$,
the number variance is reduced by about a factor two, as it is expected from
the RMT.

In order to describe the complete variation of the form factor 
or of the number variance with flux, including the $\phi_0/2$
periodicity, the expression for $P(t,\varphi)$ as a discrete  
sum (\ref{Ptphi}) over winding numbers $m$ has to be 
used\cite{Altshuler91}.  Alternatively, $P(t,\varphi)$ can also 
be expanded as a sum over  the diffusion modes ${\bf q}$, 
(eq. \ref{Pclt}). In the ring geometry, the modes are quantized
by the periodic boundary conditions\cite{Altshuler91}. For a 
quasi-1D diffusion,  $q = {2 n \pi \over L}$  for the classical 
term and $q = {2  \pi \over L}(n+ 2 \varphi)$  for the 
interference term. As a result, \begin{equation}\label{modes}
P(t,\varphi)= \sum_{n} \left( e^{-4 \pi^2 n^2 E_c t} +
e^{-4 \pi^2 (n+ 2 \varphi)^2 E_c t}\right) \end{equation}  
There is an interesting duality between the mode 
(eq. \ref{modes}) and the winding number (eq. \ref{Ptphi}) expansions. The
zero-mode approximation corresponds to a continuous summation 
over the winding numbers up to infinity,  as we have done 
above. On the other hand, the $m=0$ winding number corresponds 
to an continuous summation over the modes, i.e. to an infinite 
system where the flux dependence is lost. These two 
descriptions are related by the Poisson summation formula.

For,  $E > E_c$, i.e. $t<\tau_D$, a diffusive particle has a very small
probability to encircle the flux. More and more
diffusion
 modes become important and the number variance
 increases with the energy window as a power law.
However, the flux dependence is still very small. For $E_c \gg \Delta$, it
is found to vary as:
\begin{equation}
\Sigma^2(E,\varphi) =  \Sigma^2(E,0) -
{1 \over  \pi^2}\ln(1+4{E_c \over \gamma_\Delta} \sin^2(2 \pi \varphi))
\end{equation}

One sees that in this regime, the reduction of the
number variance is much smaller than what could have been
 naively expected. This is because the contribution to the
spectral rigidity comes from short time trajectories and most of
these trajectories cannot enclose a flux.
 This is clearly seen on fig. \ref{f3:pdet}      where the flux dependence
of $P(t,\varphi)$ is small.

Finally, let us also emphasize that, although the spectrum has the
periodicity $\phi_0$, the spectral rigidity has the periodicity
$\phi_0/2$. The Hamiltonian at $\phi = \phi_0/2$ has the GOE symmetry. This
is a situation of "false time-reversal violation"\cite{Robnik86}.

 In the next section, we study the physics of the persistent currents which
is intimately connected to the evolution of the spectral rigidity in a flux.

\section{Persistent currents}  \label{sect4}

\subsection{Introduction}
Although in the 80's most of the physics of the mesoscopic systems dealt
with transport properties,  a
wide interest  recently emerged in the experimental
studies of the equilibrium properties. Among them the search for
the persistent current of an isolated mesoscopic metallic
ring pierced by a magnetic Aharonov-Bohm flux $\phi = \varphi \phi_0$.

Such a ring carries an equilibrium magnetization $M$ which is the derivative
of the free energy $F$ with respect to the  magnetic field $H$: $M =
-{\partial F \over \partial H}$.   It corresponds to a
current $I$ flowing along the ring, which is given by
\begin{equation}
I(T) = -{\partial F\over \partial \phi}= - \sum_n f_n {\partial \ep_n \over
\partial \phi} \ \ \ \ \ \ \ I(T=0) = -{\partial E\over \partial \phi}= -
\sum_{\ep_n < \ep_F}  {\partial \ep_n \over  \partial \phi} \label{I}
\end{equation}$f_n$ is the Fermi factor associated with the energy $\ep_n$.
 For non-interacting electrons, this current is simply the sum
of the flux derivatives of each individual energy levels. Introducing the
flux dependent DOS,       $\rho(\ep,\varphi)$, the current is rewritten as:
 \begin{equation} \label{Isum}
I = - {\partial \over \partial \phi} \int_0^{\ep_F} \ep \rho(\ep,\varphi)
d\ep \end{equation}
The existence of this current has been predicted long ago by
Hund\cite{Hund38}
and first calculated by Bloch and Kulik
for a clean strictly 1D
ring\cite{Bloch65,Kulik70}. In 1983,  B\"uttiker,
Imry and Landauer proposed that it
could be also observed in a metallic disordered ring\cite{Buttiker83}.
This persistent current has been observed for the first time by Levy {\it et
al.} who measured the  magnetization of $10^7$ isolated mesoscopic $Cu$
rings\cite{Levy90}. In this experiment, the observed magnetization gives only access to
the average current of each ring $\langle I \rangle$.

The estimation of this average current raises an interesting problem.
From the spectrum of fig. \ref{f2:spectra}
we see that the energy levels move upwards and
downwards with the flux in a random way. One can thus expect  single
level current $i_n = \partial \ep_n / \partial \phi$ to be zero on average:
$\langle i_n \rangle =0$ and thus $\langle I \rangle =0$. In other words,
 $\langle \rho(\ep,\phi)\rangle$ is flux independent. This is because
the average DOS involves only trajectories of zero length so that no
flux is enclosed by these trajectories. One then concludes
that there is no average persistent current (the current is actually
exponentially small\cite{Cheung89,Entin89}.
During the last years, there has been a large amount of work to explain the
origin of a non-zero average current.

On the other hand, two  experiments have been performed on
single rings, one on $Au$ rings\cite{Chandrasekhar91}
, the other  on $GaAs-GaAlAs$ rings\cite{Mailly93}.
 The
relevant quantity in these cases is the typical current
 $I_{typ} = (\langle I ^2 \rangle -  \langle I \rangle^2)^{1/2} \simeq 
(\langle I ^2 \rangle)^{1/2}$.

In the following subsection, we present the calculation of the typical and
average persistent current for diffusive non-interacting electrons.

\subsection{The typical current}
From the expression of the current in terms of the DOS, the typical
current $I_{typ}$ is
given by
\begin{equation}
 I^2_{typ} = {\partial \over \partial \phi}
{\partial \over \partial \phi'} \int_{-\ep_F}^0 \int_{-\ep_F}^0 \ep \ep'
K(\ep-\ep',\varphi,\varphi') d\ep d\ep'  \label{Itypical}
\end{equation}
where the correlation function $K$ is the sum of the classical
 and of the interference terms:
\begin{equation}  \label{decouplingK}
K(\omega,\varphi,\varphi')=   K_{cl}(\omega,{\varphi-\varphi'\over 2})+
K_{int}(\omega,{\varphi+\varphi'\over 2})
 \end{equation}
Similarly: \begin{equation}  \label{decouplingP}
P(\omega,\varphi,\varphi')=   P_{cl}(\omega,{\varphi-\varphi'\over 2})+
P_{int}(\omega,{\varphi+\varphi'\over 2})
 \end{equation}
 Fourier transforming $K(\ep-\ep')$ and using the identity
 $\int_0^{\infty} \ep d \ep e^{i\ep t} =-1/t^2$, one obtains
straightforwardly:
\begin{eqnarray}  \label{Itypical2}
\langle I^2 \rangle
&=&
{1 \over 2 \phi_0^2}
\int_0^\infty
{\tilde{K}_{int}''(t,\varphi) - \tilde{K}_{cl}''(t,0)
\over t^4} d t
\\
&=&
{1 \over 8 \pi^2 \phi_0^2}
\int_0^\infty
{P_{int}''(t,\varphi)- P_{cl}''(t,0)
\over t^3} d t
\end{eqnarray}
where $''$ denotes the second derivative $\partial^2 /\partial \varphi^2$.
The expansion of the return probability in winding
numbers gives directly the harmonics decomposition of the 
typical current: \begin{equation} \label{Itypical3} \langle I^2 
\rangle =    \sum_{m=1}^\infty  \langle I^2 \rangle_m   \sin^2(2 \pi m
\varphi) \end{equation} with \begin{equation}\label{Itypicalm} 
\langle I^2 \rangle_m = { 8 m^2 \over   \phi_0^2}
\int_0^\infty  {P_m(t)\over t^3} dt \end{equation} where 
$P_m(t)= {L \over \sqrt{4 \pi E_c t}} e^{-m^2/4 E_c t}$. We use
($\alpha \geq 1$): \begin{eqnarray}\label{integral}
\int_0^\infty  {P_m(t)\over t^\alpha} dt &=&  {2^{2 
\alpha-1}\over \sqrt{\pi}}{E_c^{\alpha-1} \over m^{2 \alpha 
-1}}\int_0^\infty w^{2 \alpha -2}e^{-w^2} d w \nonumber \\ &=& 
2^{ \alpha -1} (2 \alpha -3)!! {E_c^{\alpha-1}\over m^{2 \alpha 
-1}}. \end{eqnarray}  where the
intermediate integral has the simple meaning of an
average dimensionless winding number since $w^2 = m^2 L^2 /4 D 
t$. 
Taking into account
the exponential reduction $e^{- \gamma t}$ of the return probability
at large times due to inelastic scattering, one finds that the harmonics
are exponentially damped by a factor $e^{- m \sqrt{\gamma / E_c}}=
e^{- m L / L_\varphi}$. The 
harmonics of the typical current are then given
by: \begin{equation} \langle I^2 \rangle_m = {96
\over m^3} ({E_c \over \phi_0})^2 [1 + {m\over 2}{L \over
L_\varphi}+{m^2 \over 3} ({L \over L_\varphi})^2] e^{-m L/L_\varphi}
\end{equation} This result should be multiplied by $4$ to take the spin
degeneracy into account, so that the typical current is multiplied by $2$.
In the limit where $L \ll L_\varphi$, the first term in brackets
gives the main contribution\cite{Argaman93,Altshuler87}. It is quite easy to
understand qualitatively why the typical current
is proportional to $E_c$ (This proportionality has been first derived in
ref.\cite{Cheung89}. In that calculation, a two-cooperon diagram is
missing). It is of the order
of the charge $e$ divided by the  characteristic time which is nothing but
$\tau_D$: $I_{typ} \propto e/\tau_D \propto E_c /\phi_0 \propto 
{e v_F \over L}{l_e \over L}$.
It turns out that the calculation of this typical current cannot
explain the amplitude observed in the experiment performed on 
single $Au$ rings\cite{Chandrasekhar91}. The order of magnitude 
of the observed current is closer to ${e v_F \over L}$.
At the moment, there is no theoretical explanation for this
discrepancy\cite{Kirczenow95}.
 The experiment performed on $GaAs-GaAlAs$ rings corresponds
to a situation where $l_e \simeq L$ and the observed current corresponds
 to the theoretical prediction.

 It is also interesting to calculate the single level typical current, 
$i_{typ}$. Similarly to the typical total current,  it is given by
\begin{equation}
 i^2_{typ} = \Delta^2 {\partial \over \partial \phi}
{\partial \over \partial \phi'} \int_0^{\ep_F} \int_0^{\ep_F}
K(\ep-\ep',\varphi,\varphi') d\ep d\ep'  \label{itypical}
\end{equation}
It differs from the typical total current only by two energy factors.
Therefore, this implies an additional $t^2$ term in the integral on time.
One finds easily:  \begin{equation}\label{itypicalm}
\langle i^2 \rangle_m =
8 m^2 ({\Delta \over   \phi_0})^2 \int_0^\infty  {P_m(t)\over t} dt
\end{equation}
From the general expression of the integral \ref{integral}, we obtain
\begin{equation}\label{itypicalm2}
\langle i^2 \rangle_m =
8 m ({\Delta \over   \phi_0})^2  e^{-m L/L\varphi}
\end{equation}
This  expansion in harmonics diverges for $L_\varphi \rightarrow \infty$.
This is because the semi-classical
calculation breaks down for long times $t \geq \tau_H$, i.e. for large
winding  numbers $m \geq \sqrt{\Delta/ E_c}$.  It turns
out that this expression of the typical single level current is 
only approximate and does not describe properly the current
resulting from numerical simulations\cite{Braun94}. This is because, in the
calculation of $\langle i^2 \rangle$, all energy scales are smaller than
$\Delta$. A calculation
based on the supersymmetry method is needed\cite{Taniguchi94}.

Finally, it is interesting to notice that a given harmonics of the typical 
total current
is $g=E_c/\Delta$ times larger than the single level current. This means that
although the total current is the sum of the currents of 
all the energy levels
in the spectrum, the contribution of the levels nearly cancel and only the 
last $E_c / \Delta$
levels contribute to the total current.
This is seen more precisely with a calculation of the correlation between
the harmonics of the current taken at different energies:
similarly to eq. \ref{itypicalm}, one finds
\begin{eqnarray}\label{itypicalEm}
\langle i_E i_{E+\omega} \rangle_m &=&
8 m^2 \left({\Delta \over   \phi_0}\right)^2 \int_0^\infty  {P_m(t)\over t}e^{i \omega t}
dt \nonumber \\
&=& 8 m^2 \left({\Delta \over   \phi_0}\right)^2 \Re[e^{-m\sqrt{{\gamma + i \omega \over
E_c}}}] \end{eqnarray}
where $i_E$ is the current at energy $E$. One sees that this correlation
function decreases on a scale $E_c /m^2$\cite{Gefen92}.

\subsection{The average current}
We turn now to the calculation of the average current relevant for
 the many rings experiment\cite{Levy90}.
\subsubsection{The canonical current}

The above calculation of the average persistent current, $\langle I
\rangle =0$,
implicitly assumes that the Fermi level is flux independent, so that in
eq. \ref{Isum}, the only flux dependence is contained in the DOS.
It has been proposed that the fact that in each ring, the
number of particles $N$ is fixed and not the Fermi level, plays an
important role and leads to a finite average current\cite{Bouchiat89}. The
reason is that the constraint that $N=\int_0^{\ep_F}
\rho(\ep,\phi)
d\ep$ is fixed implies that $\ep_F$ must be flux dependent. This flux
dependence in eq. \ref{Isum} being correlated to the flux dependence of
$\rho(\ep,\phi)$ makes the average current non zero \cite{Imry91}. This
"canonical"\footnote{The word canonical is here slightly 
misleading. This calculation is actually performed in the grand-canonical 
ensemble, with  a constraint that the average number of 
particles is fixed and flux independent}
average can be found by doing an expansion around the average Fermi level
$\ep_F$. Denoting by $\mu(\phi)$ the sample and flux dependent chemical
potential and $\ep_F = \langle \mu(\phi)\rangle$ one finds:
 \begin{eqnarray}
I_N &=& -{\partial F \over \partial \phi}|_N = -{\partial \Omega \over
\partial \phi}|_{\mu(\phi)}  \\
 &=& -{\partial \Omega \over
\partial \phi}|_{\ep_F} - {\partial^2 \Omega \over
\partial \mu \partial \phi}|_{\ep_F}( \mu(\phi) - {\ep_F})
 \end{eqnarray}
By definition, the first term of this expression is the grand canonical
current $I_{\mu}$. The first derivative $-{\partial \Omega
\over \partial \mu}$ is the number of particles  $N$. As a result:
\begin{eqnarray}
I_N &=& I_{\ep_F}+{\partial N \over
\partial \phi}|_{\ep_F} ( \mu(\phi) - {\ep_F})
 \end{eqnarray}
We then use the relation $\delta \mu |_N = -\Delta \delta N |_{\mu}$ which
expresses that the variation of the chemical potential at fixed number of
particles is proportional to the variation of the number of particles at
fixed chemical potential\cite{cautious}. After averaging,  neglecting
 $\langle I_{\ep_F} \rangle$, one deduces the following
relation\cite{Imry91}:
\begin{equation}  \label{IN}
\langle I_N \rangle =  -{\Delta
\over 2} {\partial
 \over \partial \phi}\langle \delta N^2({\ep_F},\phi)\rangle =
 -{\Delta \over 2} {\partial
 \over \partial \phi}
\int_0^{\ep_F}\int_0^{\ep_F} K(\ep,\ep ',\phi) d \ep d\ep'
 \end{equation}
Therefore, when the number of electrons in the rings is fixed,
the average persistent current is finite and is
rewritten
in terms of the typical  sample
to sample fluctuation in the number of levels below the Fermi
energy $\ep_F$\cite{Altshuler91,Schmid91,VonOppen91,Akkermans91}. From
the above semi-classical analysis, the current can be directly written in
terms of $P(t)$: \begin{equation}    \label{Iaver}
\langle I_N \rangle =-{ \Delta }  {\partial \over \partial \phi}
\int_0^\infty  {\tilde{K}(t,\varphi) \over t^2}  d t = -{ \Delta \over 4
\pi ^2} {\partial \over \partial \phi}
\int_0^\infty  {P(t,\varphi) \over t} d t\end{equation}
The integral is known (eq. \ref{integral}) and gives the harmonics expansion
\begin{equation}    \label{Iaver2}
\langle I_N \rangle= \sum_{m=1}^\infty \langle I_N \rangle_m \sin(4 \pi m
\varphi) \end{equation}
with
\begin{equation}    \label{Iaver3}
\langle I_N \rangle_m={2 \over \pi}{\Delta  \over \phi_0}e^{- m L/L_\varphi}
\end{equation}
The average current can be reconstructed as\cite{Oh91}
\begin{equation}    \label{Iaver4}
\langle I_N \rangle={\Delta  \over \pi \phi_0}{\sin 4 \pi  \varphi
\over \cosh L/L_\varphi - \cos  4 \pi  \varphi}\end{equation}
In order to take into account the spin of the electrons, the number
variance should be multiplied by $4$  and the interlevel spacing divided by
$2$ so that the current \ref{Iaver4} is doubled.

This current oscillates with the period $\phi_0/2$ and is {\it
paramagnetic} at small flux. This is clearly seen from eq. \ref{IN}: the
average current measures the change in spectral rigidity when the TRS is
broken. The number variance decreases when $\phi$ is finite so that the
current has to be paramagnetic.
The order of magnitude of this average "canonical" current is smaller than 
observed experimentally for an ensemble of $10^7$ $Cu$ rings\cite{Levy90}.

For completeness, we mention here that the magnetism of small disordered
metallic dots -- where the external parameter is a magnetic flux in the bulk
of the dot instead of an AB flux -- can be also described along
the same ideas\cite{magnetization}.

 \subsubsection{electron-electron
interactions}
Another larger contribution to the persistent current may come from interactions
as  has been first proposed by Ambegaokar and Eckern, in the
framework of
the Hartree-Fock(HF) approximation and using diagrammatic 
calculations\cite{Ambegaokar90,Schmid91}. We present here a simple 
semi-classical derivation of this contribution to the 
persistent current\cite{Montambaux95}. A different approach, based on
Density
Functional Theory, leads to similar results\cite{Argaman94}. The HF
equations read, for each state $i$ \begin{eqnarray}
\ep_i \psi_i(\r) & = &
 T \psi_i(\r) + V(\r) \psi_i(\r) +
\sum_j \int U(\r-\r') | \psi_j(\r')  |^2 \psi_i(\r) d\r' \nonumber \\
& & - \sum_j
\delta_{\sigma_i \sigma_j} \int U(\r-\r')  \psi_j^{\ast}(\r') \psi_j(\r)
\psi_i(\r') d\r'
\label{hf1}
\end{eqnarray}

In perturbation to the first order in the interaction parameter $U$, the
shift of the energy levels is thus given by
 \begin{eqnarray}
\ep_i & = & \ep_i^0 +
\sum_j \int U(\r-\r') |\psi_j(\r')|^2 |\psi_i(\r)|^2  d\r d\r'   \nonumber  \\
&  & - \sum_j
\delta_{\sigma_i \sigma_j} \int U(\r-\r')  \psi_j^{\ast}(\r') \psi_j(\r)
\psi_i^{\ast}(\r) \psi_i(\r') d\r d\r'
\label{hf2}
\end{eqnarray}
\noindent where the states $\psi_i$ are those of the
{\it non-interacting} system. As a result, the total energy $E_T$ is now
\begin{eqnarray} E_T & = & E_T^0 +
 {1 \over 2}\sum_{i,j} \int U(\r-\r') |\psi_j(\r')|^2 |\psi_i(\r)|^2  d\r
d\r' \nonumber  \\ & & - {1 \over 2} \sum_{i,j}
\delta_{\sigma_i \sigma_j} \int U(\r-\r')  \psi_j^{\ast}(\r') \psi_j(\r)
\psi_i^{\ast}(\r) \psi_i(\r') d\r d\r'
\label{hf3}
\end{eqnarray}
\noindent where $E_T^0$ is the total energy in the absence of interaction.
The summation $\sum_{i,j}$ is  on filled energy levels. $\sigma_i$ is
the spin of a state $\psi_i$. In terms of the local
density $n(\r) = \sum_{i} |\psi_i(\r)|^2$, the total energy can be rewritten:
\begin{eqnarray}
E_T & = & E_T^0 + {1 \over 2}
\int U(\r-\r') n(\r') n(\r)  d\r d\r'    \nonumber  \\
&  & - {1 \over 2}\sum_{i,j}
\delta_{\sigma_i \sigma_j} \int U(\r-\r')  \psi_j^{\ast}(\r') \psi_j(\r)
\psi_i^{\ast}(\r) \psi_i(\r') d\r d\r'
\label{hf4}
\end{eqnarray}

We now assume a  screened Coulomb interaction:
$U(\r-\r') = U \delta(\r-\r')$ where $U=4 \pi e^2 / q_{TF}^2$, $q_{TF}$ being
the the Thomas-Fermi wave vector. For such a local interaction, the Fock
term gets the same structure as the Hartree term and one
has:
\begin{equation} E_T = E_T^0 +  {U  \over 2} \int  n^2(\r)  d\r
-
{U \over 4}  \int n^2(\r) d\r
\label{hfscr}
\end{equation}

\noindent Because of the spin,  the Hartree contribution is
twice the Fock contribution\cite{Ambegaokar90}.
The interaction gives a new contribution to the persistent current
\begin{equation}  \label{currentee}
 \langle I_{e-e} \rangle =  -\langle{\partial E_T \over \partial 
\phi}\rangle = -{U \over
4} {\partial \over \partial \phi} \int \langle n^2(\r) \rangle
d\r
\end{equation}
We define the local DOS $\rho(\r,\omega)$ so that
$n(\r) = 2 \int_0^\mu \rho(\r,\omega) d\omega$ ( the factor $2$ accounts
for spin). The current
can be rewritten as: \begin{eqnarray} \langle I_{e-e} \rangle &=&  - U
 {\partial \over \partial \phi} \int \langle \rho(\r,\omega_1)
\rho(\r,\omega_2) \rangle d\r  d\omega_1  d\omega_2  \\
&=&  -{U
\over  2 \pi^2} {\partial \over \partial \phi} \int \langle
G^R(\r,\r,\omega_1)
G^A(\r,\r,\omega_2) \rangle d\r  d\omega_1  d\omega_2    \\
&=&  -{U \rho_0
\over  \pi} {\partial \over \partial \phi} \int P(\omega_1 -\omega_2)
  d\omega_1  d\omega_2  \\
&=&  -{U \rho_0
 \over  \pi} {\partial \over \partial \phi} \int_0^\infty  {P(t,\phi) \over
t^2}   dt
\label{currentee3}
\end{eqnarray}
Using again the integral (\ref{integral}), one finds easily the harmonics content
of the HF current:
\begin{equation}    \label{Ieehar}
\langle I_{e-e} \rangle= \sum_{m=1}^\infty \langle I_{e-e} \rangle_m \sin(4
\pi m \varphi) \end{equation}
with
\begin{equation} \label{currenteem}
\langle I_{e-e} \rangle_m = 16 {U \rho_0 \over \phi_0}{E_c \over m^2}[1 +
m {L \over L_\varphi}] e^{-m L/L\varphi} \end{equation}
In the limit $L \ll L_\varphi$, the second term in brackets can be
neglected and one recovers the result obtained for the
first time
with a diagrammatic
calculation\cite{Ambegaokar90,Schmid91,Altshuler85b}.
Starting from eq. \ref{currentee3}, one can actually recover the
expression resulting from that calculation\begin{equation}
\langle I_{e-e} \rangle  = {U \rho_0 \over \pi}{\partial \over
\partial \phi}
\sum_q \int_0^{\infty} \Re {-\omega   \over  \gamma + D q^2 -i \omega}
\ d\omega \label{currentee4}
\end{equation}
\noindent The quantization of the wave vectors of the diffusion modes
in now flux dependent $q(\phi)$. A Poisson summation leads to the
above Fourier expansion.

Eq. \ref{currenteem} shows that the persistent current is
proportional
to the interaction parameter. It is known that this current is smaller than
experimentally observed \cite{Levy90}. Moreover, inclusion of
higher order contributions in the interaction parameter (the
so-called Cooper
channel renormalization) reduces further the amplitude of the estimated
 current  \cite{Eckern91,Altshuler83,Smith92}. It is one order of
magnitude smaller than the observed current\cite{Levy90}.

\subsubsection{Local versus global fluctuations of the DOS}

 It is interesting to contrast the two expressions for the
canonical and Hartree-Fock average currents. The canonical current results
from a constraint on the conservation of the total number of 
particles $N$. From this constraint, the current has been
 rewritten in terms of the flux variation
of the two-point correlation function  of the {\it global} density of
states.
On the other hand,   the HF average current has been written is terms of
the two-point correlation function of the {\it local} DOS. This must reflect the
constraint
of a {\it local} conservation of a number of particles\cite{Schmid91}.
 This can be
simply understood, using a simple argument due to
Argaman and Imry\cite{Argaman94}.

Suppose that the sample can be divided into pieces $i$ in which the 
number of particles $N_i$ is fixed, due to electrostatic interaction. By
definition,
$N=\sum_i N_i$ and the energies are extensive:  $F=\sum_i F_i$ and
$\Omega=\sum_i \Omega_i$.  We can calculate the current in each box $i$ as
we did above for the canonical current.
The current can thus be written as
 \begin{eqnarray}
I_{e-e} &=& -\sum_i{\partial F_i \over \partial \phi}|_{N_i} =
 -\sum_i{\partial \Omega_i \over
\partial \phi}|_{\mu_i(\phi)}  \\
 &=& -{\partial \Omega \over
\partial \phi}|_{\ep_F} - \sum_i{\partial^2 \Omega_i \over
\partial \mu_i \partial \phi}|_{\ep_F}( \mu_i(\phi) - {\ep_F})
 \end{eqnarray}
In each box, the local chemical potential $\mu_i$ has to adjust with the
flux to keep $N_i$ constant.

 The number of particles in one box is $N_i = -{\partial \Omega_i
\over \partial \mu_i}$. Then:
\begin{eqnarray}
I_{e-e} &=& I_{\ep_F}+\sum_i{\partial N_i \over
\partial \phi}|_{\ep_F} ( \mu_i(\phi) - {\ep_F})
 \end{eqnarray}
Using  the relation $\delta \mu_i |_{N_i} = -\Delta \delta N_i 
|_{\mu_i}$ , averaging and
neglecting $\langle I_{\ep_F} \rangle$, we obtain:
\begin{equation}
\langle I_{e-e} \rangle =  -{\Delta
\over 2}\sum_i {\partial
 \over \partial \phi}\langle \delta N_i^2({\ep_F},\phi)\rangle
 \end{equation}
 to be contrasted with the canonical current:
\begin{equation}
\langle I_N \rangle =  -{\Delta
\over 2} {\partial
 \over \partial \phi}\langle \delta N^2({\ep_F},\phi)\rangle \ 
\ \ \ \ , \ \ \ \ N= \sum_i N_i
 \end{equation}
Replacing the discrete sum over the boxes by an integral, one finds (the
interaction parameter is of order $1$): \begin{equation} \langle I_{e-e}
\rangle
= -{\Delta  V \over 2} {\partial \over \partial \phi} \int \langle n^2(r)
\rangle d\r
\end{equation}
for the current resulting from a {\it local} constraint. This
current has exactly the structure of the Hartree-Fock current 
(with a prefactor describing the strength of the interaction).
On the other hand, the canonical current is
\begin{equation} \langle I_{N} \rangle =  -{\Delta
\over 2} {\partial \over \partial \phi} \int \langle n(\r) n(\r')
\rangle d\r d\r'
\end{equation}
and results from a {\it global} constraint.

It is then clear from this viewpoint that the canonical current
and the Hartree-Fock current have very similar physical 
origins, one resulting from a global conservation of the 
particle number while the other results from a fixed local
density due to the electrostatic interaction\cite{Schmid91}.

\subsection{Temperature dependence}

Up to now the physical quantities have been estimated at zero temperature.
The extension to finite $T$ can be easily done by using the identity:
$\int f(\ep) g(\ep) d\ep =
2 i \pi T \sum_{\omega_n} g(i\omega_n)$ where $f(\ep)$ is the Fermi
distribution  and $omega_n = (2 n +1) \pi T$.
The quantities of interest (except the typical current) have the form
\begin{eqnarray}
\int dE \int dE' \Gamma(E-E')
 f(E) f(E')&=& 4 \pi^2 T^2 \sum_{\omega_n}\sum_{\omega_n'}  \Gamma(i
\omega_n-i \omega_n')\\
 &=& \int  {\pi^2 T^2 \over ( \sinh \pi T t )^2} \tilde{\Gamma}(t) dt
 \end{eqnarray}
so that  the $t$-dependent integrands  involved in the
different expressions of the currents should now contain the factor
$({\pi T t \over \sinh \pi T t })^2$.
Since the characteristic time associated with winding number $m$ is $\tau_m
= m^2 / E_c$, it is clear that the temperature dependence  of the
$m^{th}$ harmonics of the current is characterized by the energy scale
$T_m=E_c /m^2$. This is consistent with the correlation energy which has been
found in eq. \ref{itypicalEm}, between the harmonics of different single level currents.

 \section{Conductance and spectrum}    \label{sect5}

The conductance of a disordered system can be derived from the Kubo
formula (see below). Its calculation requires the knowledge of both the
wave functions and  the energy levels. We would like to know whether the
conductance  could be also derived  {\it only} from the   knowledge of
the spectrum and  of
its correlations.
 For example,  the average DOS does not carry any information on the 
transport or on the degree of localization.
But we have
already seen that the conductance  can in principle be obtained from the
 spectral analysis since $E_c = g \Delta $ is the characteristic energy
scale at which the RMT correlations disappear.
We have also seen that $E_c = g \Delta $ drives the flux
sensitivity of the levels and that a thermodynamic quantity 
like the typical persistent current is
proportional to the conductance $g$. Therefore the conductance can somehow
be related to the two point-correlation function and its
variation with an external parameter like an
 AB flux.
A  simple relation has been first proposed in the 70's by Thouless, which relates the
conductance to the curvature of the energy levels. 
We analyze now this relation and more recent developments.

\subsection{The Thouless formula and its extensions}

A very important step linking the transport to the spectral properties
has been put forward by Edwards and Thouless\cite{Edwards72,Thouless74}
who argued that the electrical conductance
can be related to the sensitivity of the energy
spectrum to a {\it change in the boundary conditions}.
 Intuitively
the more a level is localized, the less sensitive it is to the boundary
conditions. Therefore the conductance should be directly related to a
measurement
of the sensitivity to the boundary conditions  (we have seen that this is
actually the case since the typical
current is proportional to the conductance). Thouless noticed the  very
similar structure between the expressions of the conductance and of
the curvature of energy levels when a  change in the boundary
conditions is introduced
$\psi(x+L)=\psi(x)e^{i\eta}$, or, equivalently, an  AB
flux: $\eta = 2 \pi \varphi = 2 \pi \phi / \phi_0$.
On the one hand, the d.c. $T=0K$ conductivity  $\sigma$
is given by the  Kubo formula:
\begin{equation} \label{sigmakubo}
 \sigma =  {\pi   e^2 \hbar \over m^2 V} \sum_{\alpha,\beta}
 |p_{\alpha\beta}|^2 \delta (E_F - \ep_\alpha)  \delta (E_F - \ep_\beta) \
\, \end{equation}
 where ${\ep_\alpha}$ is a single energy level
 and $p_{\alpha\beta} = \langle \alpha | p| \beta \rangle$ is the
matrix element of the   momentum operator along the x- direction.
 On the other hand, the curvature of a given energy level $\alpha$
  at
the origin ($\eta = 0$) is easily found from perturbation
theory. It is given by:
\begin{equation}
c_\alpha={({{{\partial}^2}{\ep_\alpha}\over{{\partial}
{{\eta}^2}}})}_{\eta=0} ={ \hbar^2\over m L^2}
+{{{ 2\hbar}^2}\over{{m^2{L^2}}}} \sum_{\beta \neq \alpha}
{ |p_{\alpha\beta}|^2
\over \ep_\alpha-\ep_\beta}
\end{equation}
In a metallic system, i.e in the presence of moderate  disorder, this
curvature is a random quantity.
Thouless assumes first  that the matrix elements $p_{\alpha\beta}$  are
not correlated with the
energy levels $\ep_\alpha$, so that the distribution of the curvature is
roughly
the distribution of the $1/(\ep_\alpha -\ep_\beta)$. Then assuming that the
energy levels are not correlated, the distribution of the
curvatures has the Cauchy form $P(c) = (\gamma_0 / \pi)/(\gamma_0^2+c^2)$
with a width $\gamma_0$ given by
\begin{equation} \label{curvature}
\gamma_0 = {2 \pi \hbar ^2  \over m^2 L^2 }
{|p_{\alpha\beta}|^2 \over \Delta} \end{equation}
On the other hand, assuming again that $p_{\alpha\beta}$ are decorrelated
from the $\ep_\alpha$, the Kubo formula gives
for the average conductivity: \begin{equation} \label{sigmakubo2}
 \sigma =  {\pi   e^2 \hbar L^d \over m^2}
  \langle |p_{\alpha\beta}|^2  \rangle
\rho_0^2
\end{equation}
${\langle}$...${\rangle}$ represents an average  over the
disorder and the energy levels. Comparison between the equations
\ref{curvature} and \ref{sigmakubo2} gives a direct relation between the
average dimensionless conductance $g=\sigma L^{d-2}/(e^2/\hbar)$ and the
width of the distribution
of curvatures, known as Thouless relation\cite{Thouless74}: \begin{equation}
\label{gg} g = {1 \over 2} {\gamma_0 \over \Delta}
\end{equation}

 The first hypothesis is reasonable and it is correct in the framework of
RMT. On the other hand, the energy levels are strongly correlated in a
metal so that the second hypothesis does not hold.
Therefore the curvature distribution may not have the Cauchy form, see
		   sect. \ref{sect6.1}.
The conductance can also be related to another quantity
connected to the flux sensitivity of the levels: it has been proposed by
Akkermans that
the conductance could also be related to the averaged square slope of the
energy
levels instead of their curvature\cite{Akkermans92a}. The conductance is
obtained as \begin{equation}
\label{gd} g_d =  \overline{{\langle i^2 \rangle \over \Delta^2}}
\end{equation}
where $i(\eta) = {\partial \ep_\alpha \over \partial \eta}$.
$\overline{ ...}$ is an average on flux.
A similar relation has also been proposed by Wilkinson\cite{Wilkinson88}
and by Simons {\it et al.}\cite{Simons93a,Simons93b,Simons93c} in the case
where the external parameter, instead of being an AB flux, does
not break any symmetry. $g_d$ measures a global, $\varphi$-averaged,
property of the spectrum, while $g$ given by eq. \ref{gg} measures a local,
$\varphi$ ${\to} 0$, property.
It has been argued that these two quantities which are two different
measures of the sensitivity of the energy levels to the boundary conditions,
should
 actually contain the same information and should be proportional in the
diffusive regime\cite{Akkermans92b}.
This aspect will be discussed in section \ref{sect6.1}.

\subsection{Universal Conductance Fluctuations}

\begin{figure}[ht]
\centerline{
\epsfxsize 8cm
\epsffile{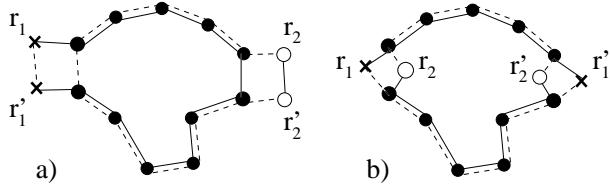}}
\caption{Schematic view of the paths contributing to conductance
fluctuations.} \label{f:ucf}
\end{figure}

An important signature of the coherent nature of quantum transport is
the phenomenon of Universal Conductance
Fluctuations\cite{Altshuler86,Lee87}.
When a physical parameter is varied, such as the Fermi energy, the magnetic
field or the disorder configuration, the conductance fluctuates around its
average value. These fluctuations are reproducible and are the signature of
the interference pattern associated to a given impurity configuration.
The width of the distribution is universal and of the order of $e^2/h$\cite{Lee87}.
 As
seen in section \ref{sect1},
the origin of this universality is  related to the spectral rigidity
of the spectrum\cite{Altshuler86}: the conductance $g$ is equal to the
number of
levels $N(E_c)$ in a strip of size $E_c$. The spectral rigidity implies that
$var[N(E_c)] \sim 1$ so that $var[g] \sim 1$. 
This variance can be calculated more precisely directly from the Kubo formula\cite{Altshuler86,Lee87}.
The averaged square of the conductance contains terms of the form
 $\langle G^R(\r_1 ,\r'_1)G^A(\r'_1 ,\r_1)
G^R(\r_2 ,\r'_2)G^A(\r'_2 ,\r_2)\rangle$ (for clarity we omitted the
gradients). As shown on fig. \ref{f:ucf}, two contractions are possible:
$\r'_1=\r_1 , \r'_2=\r_2$ and  $\r_2=\r_1 , \r'_2=\r'_1$.
The first term a) is proportional to $\int
G^R(\r_1,\r_1,t)G^A(\r_2,\r_2,t)dt
d\r_1 d\r_2$. For the same reason as in section \ref{Semi},
$\r_1$ and $\r_2$
belong to the same orbit of length $v_F t$.
Therefore integration on $\r_2$ gives a factor proportional to $v_F t$ and
the corresponding contribution to the conductance fluctuation has the form:
\begin{equation} \label{UCF}
{\langle \delta g^2 \rangle  \over \langle g \rangle^2}  \propto
\int_0^\infty t P(t) dt
\end{equation}
This term  has exactly the same structure as the two-point
 correlation function of the DOS. It describes the contribution
of the DOS fluctuations of the conductance
fluctuations\cite{Altshuler86,Argaman95}. The second term b) is proportional
to
 $\int P(\r,\r',t)P(\r',\r,\tau)dt d\tau d\r d\r'$. It
 can be also rewritten in the form $\int t P(t) dt$. It describes the
contribution of the fluctuations of the diffusion coefficient to the
conductance fluctuations\cite{Altshuler86,Argaman95}.
 The integral \ref{UCF}
scales as $\tau_D^2 \propto  1/ \langle g \rangle^2 $. One then concludes
 that the fluctuations
are universal.

\section{Parametric correlations}    \label{sect6}

\subsection{Curvatures distribution}       \label{sect6.1}

Thouless assumed that the levels are uncorrelated and found a Cauchy
distribution for the curvatures. As we know from the RMT, levels are
actually strongly correlated and repel each other. This repulsion must
affect the curvature distribution $P(c)$. It is actually easy to find the
tail of the distribution  $P(c)$. When two levels are very close in energy $s
\rightarrow 0$, one can isolate this pair and treat it in a perturbative
way. 
The distance $s(\lambda)$ where $\lambda$ is the perturbation 
parameter, varies as $\sqrt{s^2+ \lambda^2} \sim s
+\lambda^2/2s$
so that for $s \rightarrow 0$, the curvature $c$ varies as $1/s$. Since
$P(s)
\propto s^\beta$ one concludes that $P(c) \rightarrow 1/c^{2+\beta}$ for $c
\rightarrow \infty$, in contradiction with the Cauchy
form\cite{Gaspard90}.

\begin{figure}[ht]
\epsfxsize 5cm
\centerline{\epsffile{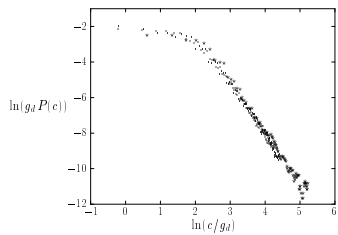}}
\caption{After rescaling the
distribution of
curvatures becomes universal.
($8\times 8\times 8$ with $w=4,5,6$)}
\label{figPcuni}
\end{figure}

The problem of the curvature distribution has been solved 
recently. In the case of pure symmetry, i.e. the perturbation parameter
$\lambda$ does not break any symmetry ( for example a step 
potential in the GOE case or an AB flux in the unitary case 
where there is already a magnetic field), the curvature 
distribution is given by: \begin{equation}\label{curvatured} 
P_\beta(c) = {N_\beta \over (\gamma_\beta^2 + 
c^2)^{(\beta+2)/2} } \end{equation} $N_\beta$ is
a normalization coefficient. This
form
has first been guessed by Zakrzewski and Delande\cite{Zakrzewski93}
to
fit numerical calculations on various model exhibiting chaotic spectra: the
kicked-top model\cite{Haake87}, random matrices, the kicked
rotator and the stadium billiard\cite{Takami92}, the hydrogen atom in a
magnetic field. The width  $\gamma_\beta$ of the
distribution is proportional to the average square of the level
velocities\cite{Gaspard90,Zakrzewski93}:
 \begin{equation}\label{gammabeta}\gamma_\beta= \pi \beta {\langle
\overline{i^2(\lambda)}\rangle\over\Delta}\end{equation}
where $i = {\partial \ep_\alpha \over \partial \lambda}$. Afterwards, this
curvature distribution has
been proven analytically by Von Oppen\cite{VonOppen94} for random matrices
of the form $H(\lambda) = H + \lambda K$ where  $H$ and $K$ are
random matrices belonging to the same symmetry and $\lambda$ is 
the perturbation parameter. Recent numerical calculations have
shown that this distribution is also characteristic of metallic 
spectra when the perturbation parameter is an AB  flux
$\phi$\cite{Braun94}(see fig. \ref{figPcuni}).
In particular, in the limit where $\phi \rightarrow 0$, the 
distribution is still the GOE distribution ($\beta=1$ in eq. 
\ref{curvatured})\cite{Braun94}.
This has been proven analytically by Fyodorov and Sommers
who also found that
there are no corrections of order $1/g$\cite{Fyodorov95}.

In the literature, the coefficient $\gamma$ entering  the Thouless
relation is sometimes taken as the typical curvature. This cannot be the
case
because, due to the $1/c^3$ tail,
the distribution has no second moment. Instead, one can choose as a
definition of the conductance\cite{Kamenev94,Casati94,Braun94,Zyczkowski94}:
\begin{equation} g_c \equiv  { \langle | c |
\rangle\over \Delta} \end{equation}
For  pure symmetry   cases,   eq. \ref{gammabeta} leads to universal
relations between $g_c$ and $g_d$\cite{Casati94}:
\begin{eqnarray}
GOE \ \ \ \ \langle | c | \rangle
= \gamma_1 \ \ \ \ and \ \ \ \gamma_1 = \pi {\overline{\langle i^2(\lambda)
\rangle} \over \Delta} \ \  &\Rightarrow&  \ \ \ g_c = \pi g_d
\label{gcgd1} \\
GUE \ \ \langle | c | \rangle
= {2  \over \pi}\gamma_2  \ \ and \ \ \gamma_2 = 2 \pi {\overline{\langle
i^2(\lambda) \rangle} \over \Delta} \ \  &\Rightarrow& \ \ \ g_c = 4
g_d \label{gcgd2}
\end{eqnarray}
 The case of an AB flux has been studied
recently numerically\cite{Braun94} and analytically with the
supersymmetry method\cite{Fyodorov95}. The result is that
\begin{equation}\label{gcgd} g_c = 2 \pi g_d \end{equation}

This relation can be simply understood:
the distribution of curvatures in zero flux is characteristic of the
GOE
symmetry so that  $\langle | c | \rangle$ is equal to the width $\gamma$ of
the GOE ($\beta=1$) distribution (eq. \ref{gcgd1}). But there is a finite
current only when $\varphi \neq 0$, i.e. when the
the symmetry has become GUE, so that the relation between $\gamma$ and
 $\langle \overline{i^2(\eta)} \rangle$ is the relation of the GUE
($\beta=2$)
symmetry (eq. \ref{gcgd2}). Combining these two relations, one gets
$  g_c = 2 \pi g_d  $.

 When $\varphi \gg \varphi_c$, the
curvature distribution
becomes the one of the unitary case with a $1/c^4$ tail, and the second
moment now converges.
In the cross-over regime, it has been found numerically that 
the curvature distribution is quite different than in the pure 
cases, with a Gaussian tail. The typical curvature in the 
cross-over regime has been found to diverge logarithmically at small flux
\cite{Braun94,Kamenev94}:
\begin{equation}
\langle c^2(\varphi) \rangle \propto   g^2 \ln (1/ g \varphi^2)
\end{equation}

\subsection{Parametric correlations}

We have seen in section \ref{sect3} that the transition between the GOE and
GUE symmetries in the diffusive regime is a universal function of the
combination of parameters $E_c \varphi^2$ or $g \varphi^2$.

Another quantity has been recently introduced by Altshuler {\it et al.}
 to characterize the motion of the  energy
levels\cite{Szafer93,Simons93a}.
It is the autocorrelation of the current for a given level. It is
defined
as  $C(\varphi_-) = {1 \over \Delta^2} \langle \overline {i(\varphi)
i(\varphi+\varphi_-)} \rangle$, where $\overline{...}$ is an average on
flux. Since the flux drives a transition between
different symmetries, the product $\langle {i(\varphi)i(\varphi+\varphi_-)}
\rangle$ depends on both $\varphi$ and $\varphi_-$. It would be translation
invariant (independent of $\varphi$) in the case where the parameter
$\varphi$ does
not break any symmetry. Here, one has to average on $\varphi$ so that the
correlation function is only a function of $\varphi_-$.
One can use the same method as in the previous sections to calculate this
correlation function.
From the definition of the current and following the same lines as for the
calculation of the typical current, one finds\cite{Szafer93}
\begin{equation}C(\varphi_-) = - {\partial^2 \over \partial
\varphi_-^2}  \int \int
 K_{cl}(\ep,\ep',{\varphi_- \over 2}) d \ep d \ep'
\end{equation}
By Fourier transform and using the relation between $\tilde K(t)$ and
$P(t)$, one deduces\cite{Goldberg91}:
\begin{eqnarray}
C(\varphi_-) &=& -
2 {\partial^2 \over \partial \varphi_-^2} \int
 {\tilde K(t,\varphi_-/ 2) \over  t^2 } dt
\nonumber \\ &=& -{1 \over 2 \pi^2}  {\partial^2 \over \partial \varphi_-^2}
 \int
 {P(t,\varphi_- / 2) \over t} d t
\end{eqnarray}
For $\varphi_- \ll 1$, $P(t)$ has the diffusive form: $P(t,\varphi / 2) =
e^{-4 \pi^2 E_c \varphi_-^2 t - \gamma t}$, so that
\begin{equation}
C(\varphi_-) = C(0) {1 - b \varphi_-^2\over (1  + b \varphi_-^2)^2}
\end{equation}
where $b= \pi^2 C(0)$ and $C(0) ={\overline {\langle i^2(\varphi) \rangle}
}= 4 E_c / \gamma$ is the average single level typical current.
For large flux $\varphi_- \gg \varphi_c=1/\sqrt{g}$, the correlation
function has a universal tail\cite{Szafer93,Beenakker93}:
\begin{equation}
C(\varphi_-) =  -{ 1  \over \pi^2 \varphi_-^2}
\end{equation}
\begin{figure}[ht]
\epsfxsize 5cm
\centerline{\epsffile{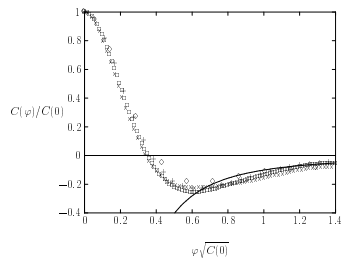}}
\caption{Universal function
$C(\varphi_-)$}
\label{figCphi}
\end{figure}
This form has also been found by Berry and Keating in the case of
billiards\cite{Berry94}. It has the same origin, it is related to the
diffusive accumulation of the phase, even when the motion is ballistic.

The expression found for  $C(\varphi_-)$ does not fit correctly the
numerical
result for small flux. The reason is that the semi-classical calculation
holds when at least one energy scale is larger than $\Delta$, which
implies here $E_c \varphi^2 > \Delta$. Thus the semi-classical calculation
describes only the power-law tail of  $C(\varphi_-)$, when $\varphi >
\varphi_c$. On the other hand, it
 has been show that the small flux behavior of $C(\varphi_-)$ has a
logarithmic correction:  $C(\varphi_-)/C(0) - 1
\propto g \varphi^2 \ln 1/ g \varphi^2
$\cite{Kamenev94,Braun94,Taniguchi94,Guarneri94}.
This is reminiscent of the logarithmic divergence of the typical
curvature near the GOE point $\varphi=0$.

As the GOE-GUE transition is driven by the unique combination of parameters
$E_c \varphi^2$, it is seen that the parametric correlation function
$C(\varphi_-)/C(0)$ is a universal function of $E_c \varphi_-^2$. This
universality has been stressed by Simons {\it et al.}
who calculated several other several parametric functions, using the
supersymmetric method\cite{Simons93a,Simons93b,Simons93c}.
 \section{Conclusion}
In this presentation of the spectral correlations and or their dependence
on the Aharonov-Bohm flux, we have tried to unify several
spectral quantities by relating all of them to
the same  quantity $P(t)$ , the return probability
for a  diffusive particle. Here we summarize schematically
the structure of
several of these quantities, together with the weak-localization correction
and with the universal conductance fluctuations.
This  description is based on a semi-classical picture in the diagonal
approximation and  fails when all energy scales become smaller
than the interlevel spacing.
 \bigskip

\centerline{
\begin{tabular}{|llll|}  \hline
 & & & \\
$K(\epsilon)$
& $\rightarrow$
&
& $\int t P(t) e^{i \epsilon t} dt$ \\
       & & & \\
$\Sigma^2(E)$
& $\rightarrow$
&
& $\int {P(t) \over t} \sin^2({E t\over 2}) dt$ \\
& & & \\
$\langle I^2 \rangle$
& $\rightarrow$
& ${\partial^2 \over \partial \varphi^2}$
& $\int {P(t) \over t^3} dt$ \\
 & & & \\
$\langle I_{e-e} \rangle$
& $\rightarrow$
& ${\partial \over \partial \varphi}$
& $\int {P(t) \over t^2} dt$ \\
 & & & \\
$\langle i^2 \rangle$
& $\rightarrow$
& ${\partial^2 \over \partial \varphi^2}$
& $\int {P(t) \over t} dt$ \\
 & & & \\
$\langle I_N \rangle$
& $\rightarrow$
& ${\partial \over \partial \varphi}$
& $\int {P(t) \over t} dt$ \\
 & & & \\
$\langle \Delta G \rangle$
& $\rightarrow$
& $$
& $\int P(t)  dt$ \\
 & & & \\
$\langle \delta G^2 \rangle$
& $\rightarrow$
& $$
& $\int t P(t)  dt$ \\
 & & & \\
\hline
\end{tabular} }
\bigskip

We have recovered in a simple way  the main quantities describing
the persistent
currents  and the parametric correlations.
 In these quantities, the flux dependence appears through the
combination  $g \varphi^2$, where $g$ is the dimensionless
conductance. This combination is the typical phase accumulated by a
diffusive particle during the Heisenberg time $\tau_H$. The observed order
of magnitude of the
persistent current  in the experiments is still unexplained. A promising way
could be a better understanding of the role of e--e
interaction. Several recent works have explored this
direction\cite{interactions}.

In this course, we have restricted ourselves to the description of
level correlations in the diffusive regime.
There has also been a recent activity on the  study of the correlations
at the Metal-Insulator transition which occurs
in 3D. In the strongly  insulating regime, the wave functions of two states
which are close in energy do not overlap. The statistics becomes Poissonian.
The Metal-Insulator transition is characterized by a third distribution,
intermediate between Wigner and Poisson distributions\cite{mit}.

Finally, the thermodynamic properties such as the magnetization of
ballistic
 mesoscopic conductors, where the disorder is so weak that
$l_e > L$ raises new interesting questions.
\section*{Acknowledgments}
A careful reading of the manuscript as well as many discussions with
 E. Akkermans, D. Braun and H. Bouchiat are gratefully acknowledged.


\begin{thebibliography}{99}
\bibitem{diffusion} This diffusive expression can be deduced directly from
the Schr\"odinger equation, doing a perturbative expansion in disorder, for
$t \ll \tau_H$ and $k_F l_e \gg 1$. See for
example \cite{Chakravarty86,Bergmann84}.
\end{thebibliography}
\end{document}